\documentclass[a4paper,11pt]{article}
\usepackage{jheppub} 
\usepackage{amsmath}
\usepackage{amssymb}
\usepackage{amsthm}
\usepackage{comment} 
\usepackage{tikz}
\usetikzlibrary{angles, quotes}
\usepackage{csquotes}
\usepackage{physics}
\usepackage[normalem]{ulem}
\newcommand{\be}{\begin{equation}}
\newcommand{\ee}{\end{equation}}
\newcommand{\hgInt}{\mathcal{F}_{4pt}}

\newcommand{\hgCoeff}{Y}
\newcommand{\threeP}{V}
\newcommand{\SthreeP}{\mathcal{V}}

\title{\boldmath Shadow formalism for supersymmetric conformal blocks}

\author{V. Belavin, J. Ramos Cabezas, B. Runov}
\affiliation{Physics Department, Ariel University, 
Ariel 40700, Israel.}

\emailAdd{vladimirbe@ariel.ac.il, juanjose.ramoscab@msmail.ariel.ac.il, borisru@ariel.ac.il}

\abstract{Shadow formalism is a technique in two-dimensional CFT allowing straightforward computation of conformal blocks in the limit of infinitely large central charge. We generalize the construction of shadow operator for superconformal field theories. We demonstrate that shadow formalism yields known expressions for the  large-c limit of the four-point superconformal block on a plane and of the one-point superconformal block on a torus. We also explicitly find the two-point global torus superconformal block in the necklace channel and check it against the Casimir differential equation.}

\begin{document}
\maketitle
\flushbottom
\section{Introduction}

A conformal field theory (CFT) in two dimensions is greatly simplified by the presence of extended symmetry w.r.t. local conformal transformations\cite{BPZ}. This symmetry allows to express any correlation function in terms of three-point structure constants and a set of functions called conformal blocks. The latter depend on conformally invariant cross-ratios of coordinates, the central charge of the theory and conformal dimensions of the fields involved, but do not depend on the three-point constants. Knowledge of four-point blocks leads to a complete solution of the theory via bootstrap equations~\cite{Polyakov1974}. However, while the series expansion of conformal block can be computed term by term, the rapid growth of complexity of the computation makes it impractical beyond the first few terms. Several alternative approaches, such as recursion representation \cite{Zamolodchikov1984,Zamolodchikov1987}, have been investigated. The full four-point conformal block is not known for a general CFT with an arbitrary choice of dimensions. Higher-point conformal blocks, as well as conformal blocks on torus~\cite{Hadasz2010,Alkalaev2022} and Riemann surfaces of higher genus, are also of significant interest.

In the large central charge limit, the Virasoro conformal blocks reduce to so-called global conformal blocks~\cite{Cho2019}. They are analogous to the conformal blocks in higher dimensions, as only descendants of intermediate field generated by global conformal subalgebra contribute to the block in this limit. Global conformal blocks have been extensively studied both on a sphere and on a torus \cite{Alkalaev:2016fok, Alkalaev2016,Alkalaev2016b,Alkalaev2017b,Alkalaev2018,Pavlov2023,Alkalaev2019,Belavin2017,Fitzpatric2015}. It turns out that they are relevant in the holographic context~\cite{Hijano2015,Alkalaev2015,Alkalaev2015b,Kraus2017}. In particular, as it was shown in~\cite{Hijano2016,Chen2017,Dyer2017}, they compute geodesic Witten diagrams in $AdS_3$.
The shadow formalism was originally proposed in \cite{Ferrara1972,Ferrara1972b,Ferrara1973} to compute conformal blocks of scalar fields in a CFT in dimension greater than two and was subsequently generalized to fields with spin in \cite{SimonsDuffin2014}. For a 2D CFT it was demonstrated that the shadow formalism can be used to compute global conformal blocks \cite{Rosenhaus2019,Alkalaev2023}. It was also successfully applied to CFTs based on W-algebras \cite{Fateev:2011qa, Belavin2024} and Galilean CFT \cite{Chen2023}. 

The key element of the shadow formalism is a shadow operator, which is, in Virasoro case, a quasi-primary composite field of dimension $1-h$, which can be constructed for any conformal primary of dimension $h$. The key property of the shadow operator is that its two-point function with the corresponding primary field is a two-dimensional delta function.
This property allows (in the large central charge limit) to construct explicitly the projector from the Hilbert space of the theory onto the highest weight module over subalgebra of global conformal transformations ($\mathfrak{sl}(2)$ for Virasoro case) in terms of the shadow operator. 

Supersymmetric conformal field theories (SCFTs) are a key element of superstring theory~\cite{Neveu1971,Ramon1971}, they arise also in the context of AdS/CFT duality \cite{Maldacena2001,Maldacena2001b,Maldacena2002,Gaberdiel2018,Giribet2018,Ebenhardt2018,Ebenhardt2019,Ebenhardt2019b}.
The goal of the present paper is to generalize the shadow
formalism to two-dimensional $N=1$ superconformal field theory in the Neveu-Schwarz sector.

The paper is organized as follows. In section \ref{VirReview}, we review the shadow formalism for two-dimensional conformal field theories in the Virasoro case. Section \ref{SuperSC} contains key facts about $N=1$ superconformal field theory in the Neveu-Schwarz sector. In section \ref{SuperShadow}, we introduce the supersymmetric shadow operator and construct the projector onto Verma supermodules corresponding to primary superfields. In section \ref{CBExamples}, we compute the four-point conformal block on a sphere via shadow formalism. In section \ref{sec:torussb}, we recall the definition of the torus superconformal blocks. In section \ref{sec:torusbsh1}, using the shadow formalism, we compute one- and two-point torus superconformal blocks. Our results for the four-point spherical superconformal block and one-point torus superconformal block are in agreement with known results~\cite{Belavin2007,Belavin2008, Alkalaev2018} obtained by other methods. The obtained representation for the two-point torus superconformal block is new. In section \ref{sec:casimirini}, we verify that it satisfies the required differential equations, which follow from the consideration based on the $\mathfrak{osp}(1|2)$ Casimir operator. In section \ref{sec:conclusions}, we present our conclusions and comments on further research directions.

\section{$\mathfrak{sl}(2)$ shadow formalism}
\subsection{Shadow formalism on a sphere}
\label{VirReview}
For a primary field $\mathcal{O}_h$ of dimension $h$ the corresponding shadow dual field $\tilde{\mathcal{O}}_h$ is defined as \cite{Alkalaev2023}
\be
	\tilde{\mathcal{O}}_{h,\bar{h}}(z,\bar{z})=\int d^2 w \frac{\mathcal{O}_{h,\bar{h}}(w,\bar{w})}{(z-w)^{2-2h}(\bar{z}-\bar{w})^{2-2\bar{h}}} \,.
\ee
It is a quasi-primary nonlocal field of dimension
\be \label{sl2shadowdim}
    h^{*}=1-h
\ee
It can be demonstrated that upon appropriate regularization the two-point function of the shadow field and the corresponding primary field is equal to a two-dimensional delta function, while two-point function with any other primary obviously vanishes.
\be
\label{DeltaVir}
	\langle \mathcal{O}_{h,\bar{h}}(z,\bar{z})\tilde{\mathcal{O}}_{h,\bar{h}}(w,\bar{w})\rangle=\delta^2(z-w)\,.
\ee
Therefore, the following operator
\be \label{thesl2idoper1}
    \Pi_{h}=\int d^2 w \,\mathcal{O}_h(w,\bar{w})
    \mid 0 \rangle \langle 0 \mid \tilde{\mathcal{O}}_h(w,\bar{w})
\ee
is invariant under global conformal transformations and acts as a projector onto irreducible $\mathfrak{sl}(2)$ modules, satisfying
\be
    \Pi_{h_1}\Pi_{h_2}=\delta_{h_1,h_2}\Pi_{h_1}\,.
\ee
A multipoint correlation function of the primary fields can be represented as a sum of conformal partial waves
\be
    \langle \phi_{h_1} (z_1) \phi_{h_2}(z_2) \dots \phi_{h_n}(z_n)\rangle=
    \sum_{\Delta_1,\Delta_2,\dots ,\Delta_{n-3}}
    \Psi^{h_1,\dots,h_{n}}_{\Delta_1,\dots,\Delta_{n-3}}
    (z_1,z_2,\dots,z_n)\,.
\ee
In the limit of large central charge one can decompose the identity operator as a sum of the projectors $P_h$ corresponding to primary fields of the theory.
Using this decomposition, it is straightforward to obtain an integral representation for a  conformal partial wave~\cite{Rosenhaus2019}:
\be
\begin{split}
    \Psi^{h_1,\dots,h_{n}}_{\Delta_1,\dots,\Delta_{n-3}}
    (z_1,\dots,z_n)=&
    \int\prod_{i=1}^{n-3} d^2 w_i\,
    \threeP_{h_1,h_2,\Delta_1}(z_1,\bar{z}_1,z_2,\bar{z}_2,w_1,\bar{w}_1) \\
    &\times\prod_{i=1}^{n-4} 
    \threeP_{\Delta_i^{*},h_{i+2},\Delta_{i+1}}
    (w_i,\bar{w}_i,z_{i+2},\bar{z}_{i+2},w_{i+1},\bar{w}_{i+1})\\
    &\times \threeP_{\Delta_{n-3}^{*},h_{n-1},h_n}(w_{n-3},\bar{w}_{n-3},z_{n-1},\bar{z}_{n-1},z_{n},\bar{z}_n)\,.
\end{split}
\ee
where the symbol $\threeP_{h_1,h_2,h_3}$ stands for the three-point function
\be
\begin{split}
    \threeP_{h_1,h_2,h_3}(z_1,\bar{z}_1,z_2,\bar{z}_2,z_3,\bar{z}_3)=\,&C_{h_1h_2h_3}|v_{h_1,h_2,h_3}(z_1, z_2, z_3)|^2\\
    =\,&
    \langle \phi_{h_1}(z_1,\bar{z}_1)
    \phi_{h_2}(z_2,\bar{z}_2) \phi_{h_3}(z_3,\bar{z}_3)\rangle,
\end{split}
\ee
and $v_{h_1, h_2, h_3}$ above denotes the holomorphic dependence of $V_{h_1,h_2, h_3}$, namely
\begin{equation}
    v_{h_1, h_2, h_3}(z_1, z_2, z_3)= \frac{1}{z_{12}^{h_1+h_2-h_3} z_{13}^{h_1+h_3-h_2} z_{23}^{h_2+h_3-h_1}}.
\end{equation}
\subsection{Shadow formalism on a torus}
As demonstrated in \cite{Alkalaev2023, Belavin2024}, integral representations of global torus conformal blocks can be obtained by inserting projectors $\Pi_h$, defined by (\ref{thesl2idoper1}), between primary fields within the trace expression of the $n$-point conformal blocks. The insertions of operators $\Pi_h$ are analogous to the insertions of the resolution of identity $\sum_{h} \mathbb{P}_{h}^{\mathfrak{sl}(2)}$, where $\mathbb{P}_{h}^{\mathfrak{sl}(2)}$ is the projector onto the $\mathfrak{sl}(2)$ module with the highest weight $h$, within the trace when one is considering global $\mathfrak{sl}(2)$ $n$-point torus conformal blocks in the so-called necklace channel. The main difference is that for projectors (\ref{thesl2idoper1}), the first projector is inserted between the trace and the first primary field, whereas for the insertions of the resolution of identity, these occur only between the primary fields. The procedure of inserting projectors (\ref{thesl2idoper1}) results in expressing the torus conformal blocks in terms of torus conformal partial waves. For the one- and two-point torus conformal partial waves, we have the expressions
\begin{equation}\label{torusshadow1-1}
     Y_{\Delta_1}^{h_1}(q, \bar{q},z_1, \bar{z}_1)=q^{\Delta_1} \bar{q}^{\Delta_1} \int d^{2} w\left|v_{1-\Delta_1, h_1, \Delta_1}(w, z_1, q  w)\right|^{2},
\end{equation}
\begin{equation}\label{torusshadow1}
\begin{aligned}
&Y_{\Delta_1, \Delta_2}^{h_1, h_2}\left( q, \bar{q}, z_1, \bar{z}_1,  z_2, \bar{z}_2 \right)=\\& =q^{\Delta_{1}} \bar{q}^{\Delta_{1}} \int d^{2} w_{1} d^{2} w_{2}\left|v_{1-\Delta_{1}, h_{1}, \Delta_{2}}\left(w_{1}, z_{1}, w_{2}\right)\right|^{2}\left|v_{1-\Delta_{2}, h_{2}, \Delta_{1}}\left(w_{2}, z_{2}, q w_{1}\right)\right|^{2}.
\end{aligned}
\end{equation}
Notice that in the above equations, the subscripts $1- \Delta_i$ represent the conformal dimensions of the corresponding shadow field as denoted in (\ref{sl2shadowdim}). To extract the holomorphic global $\mathfrak{sl}(2)$ one- and two-point conformal blocks, one does not need to take the full two-dimensional integrals in (\ref{torusshadow1-1}, \ref{torusshadow1}). Instead, it is sufficient to work only with the holomorphic part of the integrands and then take the integral over $w_i$ over an appropriate domain. For the global $\mathfrak{sl}(2)$ one-point torus conformal block, the integral representation reads

\begin{equation} \label{torusshadow2}
\begin{aligned}
&\mathcal{F}_{\Delta_{1}}^{h_{1}}(q)=\frac{1}{c_1} q^{\Delta_{1}} \int_{0}^{z_1} d w_1 v_{1-\Delta_{1}, h_{1}, \Delta_1}(w_1, z_1, q w_1 ),
\end{aligned}
\end{equation}
where $c_{1}$ is a normalization constant given by

\begin{equation} \label{torusshadow3}
c_{1}(h_1, \Delta_1)=\frac{\Gamma\left(2 \Delta_{1}-h_{1}\right) \Gamma\left(h_{1}\right)(-1)^{2 \Delta_{1}}\left(-z_{1}\right)^{h_{1}}}{\Gamma\left(2 \Delta_{1}\right)}.
\end{equation}
From the holomorphic part of (\ref{torusshadow1}), the integral representation for $\mathfrak{sl}(2)$ two-point torus conformal block reads
\begin{equation} \label{torusshadow4}
\begin{aligned}
 &\mathcal{F}_{\Delta_{1},\Delta_2}^{h_1, h_2} (q_{1}, z_{1}, z_{2} )=\\&=\frac{q^{\Delta_1}}{c_2(h_1,h_2, \Delta_1, \Delta_2)}\int_{\mathbf{C}_{1}} d w_{1} \int_{\mathbf{C}_{2}} d w_{2} \quad v_{1-\Delta_{1}, h_{2}, \Delta_{2}}\left(w_{1}, z_{1}, w_{2}\right) v_{1-\Delta_{2}, h_{2}, \Delta_{1}}\left(w_{2}, z_{2}, w_{1} q\right)\\ &=
 \left ( \frac{z_1^{h_2}  z_2^{h_1} (1-q)^{h_1+h_2}}{z_{12}^{h_1+h_2} (z_2-q z_1)^{h_1+h_2} }\right) \rho_1^{\Delta_1} \rho_2^{\Delta_2} F_4 \left[   \substack{\Delta_1+\Delta_2-h_1, \Delta_2+\Delta_1-h_2 \\  2\Delta_1, 2 \Delta_2} | \rho_1, \rho_2\right],
\end{aligned}
\end{equation}
where $c_2$ is also a normalization constant given by (\ref{coefficientc2}), the integration domains $\mathbf{C}_1$ and $\mathbf{C}_2$ are defined accordingly

\begin{equation} \label{torusshadow5}
\begin{array}{ll}
\mathbf{C}_{2}: & w_{2} \in\left[w_{1} q, z_2 \right] ,\\
\mathbf{C}_{1}: & w_{1} \in\left[z_{2}, z_{1}\right].
\end{array}
\end{equation}
The variables $\rho_1, \rho_2$ are given by 
\begin{equation} \label{ap:2ptgcb6}
    \rho_1 = \frac{q (z_{12})^2}{ z_1 z_2 (1-q)^2},\quad \rho_2= \frac{ (z_2-q z_1)^2}{z_1 z_2 (1-q)^2},
\end{equation}
and $F_4$ is the Appell function defined as
\begin{equation} \label{ap:2ptgcb9}
        F_4 \left[   \substack{a_1, a_2 \\  c_1, c_2} | x_1, x_2\right]=\sum_{m_1, m_2=0}^{\infty} \frac{ (a_1)_{m_1+m_2} (a_2)_{m_1+m_2}}{ (c_1)_{m_1}(c_2)_{m_2}} \frac{x_1^{m_1}}{m_1!} \frac{x_2^{m_2}}{m_2!},
    \end{equation}
where, $(a_i)_m$ stands for the Pochhammer symbol. For the $\mathfrak{osp}(1|2)$ discussion, the integral (\ref{torusshadow4}) will pay a key role. Therefore, we will review it in detail in appendix \ref{app:sl2twopoint}.

\section{Superconformal field theory}
\label{SuperSC}
In this section we list key facts about $N=1$ two-dimensional superconformal field theory in Neveu-Schwartz sector, following review \cite{ALVAREZGAUME1992171}.
The $N=1$ super-Virasoro algebra in NS sector is comprised of generators $L_{k}$ and $G_{k+\frac{1}{2}}$, obeying the following commutation relations \cite{FRIEDAN198537}
\be
    [L_m,L_n]=(m-n)L_{m+n}+\delta_{m+n,0}\frac{\hat{c}}{8}\left(m^3-m\right)\,,
\ee
\be
    [L_m,G_r]=\left(\frac{m}{2}-r\right)G_{m+r}\,,
\ee
\be
    \lbrace{G_r,G_s\rbrace}=2L_{r+s}+\frac{\hat{c}}{2}\left(r^2-\frac{1}{4}\right)\,.
\ee
It is a central extension of the algebra of generators of local superconformal transformations of $\mathbb{C}^{1|1}$ superspace. The latter can be parametrized by two real and two  Grassmann numbers.  It is natural to introduce holomorphic and antiholomorphic supercoordinates on the superspace
\be \label{supercoordinates1}
    Z=(z,\theta)\,,\quad \bar{Z}=(\bar{z},\bar{\theta})
\ee
and superderivatives 
\be
    D=\partial_{\theta}+\theta\partial_z\,,\quad
    \bar{D}=\partial_{\bar{\theta}}+\bar{\theta}\partial_{\bar{z}}
\ee
obeying
\be
    D\bar{Z}=\bar{D}Z=0\,.
\ee
A function $f(Z,\bar{Z})$ on superspace is called superanalytic if it satisfies
\be
    \bar{D}f(Z,\bar{Z})=0\,.
\ee
Superanalytic functions admit Taylor-like series expansion 
\be
\label{SuperAnalytic}
    f(Z_1)=\sum_{k=0}^{\infty}\frac{Z_{12}^{k}}{k!}\partial_2^k  \left(1+(\theta_1-\theta_2)D_2\right)f(Z_2)\,,
\ee
where the quantity $Z_{12}$ (which is the supersymmetric generalization of the difference of coordinates) depends on $Z_1,Z_2$ as
\be
\label{Z12def}
    Z_{12}=z_{12}-\theta_{12}
\ee
and 
\be
    z_{12}=z_1-z_2\;,\qquad\theta_{12}=\theta_1\theta_2\,.
\ee
Superconformal transformations 
\be
\label{superConfLocal}
    Z \mapsto \tilde{Z}=(\tilde{z}(Z),\tilde{\theta}(Z))
\ee
are defined as transformations preserving the superderivative:
\be
    D=D\tilde{\theta} \tilde{D}\,.
\ee
One can also define a superdifferential $dZ$ transforming as
\be
    d\tilde{Z}=D\tilde{\theta}dZ
\ee
under superconformal transformations (\ref{superConfLocal}). The subgroup of global superconformal transformations is isomorphic to $OSp(1|2)$ and is comprised of linear fractional transformations of the form
\be
    \tilde{z}=\frac{az+b+\alpha \theta}{cz+d\beta{\theta}}\,,\quad
    \tilde{\theta}=\frac{\bar{\alpha}z+\bar{\beta}+\bar{A}\theta}{cz+d+\beta\theta}\,,
\ee
with
\be
    \bar{\alpha}=\frac{a \beta - c\alpha}{\sqrt{ad-bc}}\,,
    \quad
    \bar{\beta}=\frac{b \beta - d\alpha}{\sqrt{ad-bc}}\,,
    \quad
    \bar{A}=\sqrt{ad-bc-3\alpha\beta}\,.
\ee
It has five independent parameters, with corresponding generators given by $L_{\pm 1},L_0,G_{\pm\frac{1}{2}}$.
The fields of the superconformal field theory are operator-valued functions on the superspace. 

Since all functions of Grassmann variables are linear, any superfield can be decomposed into a linear combination of ordinary fields as follows\footnote{In the notation of primary superfields (\ref{superfieldDecomp}), for simplicity of writing, we omit the dependence on the antiholomorphic conformal dimension $\bar{h}$.}

\be
\label{superfieldDecomp}
\Phi_h(Z,\bar{Z})=\phi_h(z,\bar{z})+\theta\psi_h(z,\bar{z})+\bar{\theta}\bar{\psi}_h(z,\bar{z})+\theta\bar{\theta}\tilde{\phi}_h(z,\bar{z})\,.
\ee
A superprimary field of dimensions $(h, \bar{h})$ is defined by the requirement that the differential
\be
    \Phi_h(Z,\bar{Z})dZ^{2h}d\bar{Z}^{2\bar{h}}
\ee
is invariant under superconformal transformations. Super Virasoro algebra contains ordinary Virasoro algebra with central charge
\be
    c=\frac{3\hat{c}}{2}
\ee
as a subalgebra. The components (\ref{superfieldDecomp}) of a superprimary field are Virasoro primaries with respective conformal dimensions $(h,\bar{h})$, $(h+\frac{1}{2},\bar{h})$,  $(h,\bar{h}+\frac{1}{2})$, $(h+\frac{1}{2},\bar{h}+\frac{1}{2})$.
As in the non-supersymmetric case, global superconformal symmetry fixes two- and three-point functions up to several constants:
\be \label{2ptsupersymmetric}
    \langle \Phi_{h_1}(Z_1,\bar{Z}_1)\Phi_{h_2}(Z_2,\bar{Z}_2)\rangle=
    \frac{\delta_{h_1,h_2}\delta_{\bar{h}_1,\bar{h}_2}}{Z_{12}^{2h_1}\bar{Z}_{12}^{2\bar{h}_1}}=
    \frac{\delta_{h_1,h_2}\delta_{\bar{h}_1,\bar{h}_2}(z_{12}+2h_1 \theta_1\theta_2)(\bar{z}_{12}+2\bar{h}_1 \bar{\theta}_1\bar{\theta}_2)}{(z_{12})^{2h+1}}\,,
\ee
\be \label{3ptsupersymmetric}
\begin{aligned}
   \mathcal{V}_{h_1, h_2, h_3}(Z_1, \bar{Z}_1; Z_2, \bar{Z}_2; Z_3, \bar{Z}_3 ) &=  \langle \Phi_{h_1}(Z_1,\bar{Z}_1)\Phi_{h_2}(Z_2,\bar{Z}_2)\Phi_{h_3}(Z_3,\bar{Z}_3)\rangle=\\&=
  \frac{C_{h_1h_2h_3}+\eta_{123}\bar{\eta}_{123}\tilde{C}_{h_1h_2h_3}}{
    Z_{12}^{\gamma_{123}}\bar{Z}_{12}^{\bar{\gamma}_{123}}Z_{13}^{\gamma_{312}}\bar{Z}_{13}^{\bar{\gamma}_{312}}Z_{23}^{\gamma_{231}}\bar{Z}^{\bar{\gamma}_{231}}
    }\,,
 \end{aligned}
\ee
where we have used shorthand notations (\ref{Z12def}), the numbers $\gamma_{ijk}$ are defined as
\be
    \gamma_{ijk}=h_{i}+h_{j}-h_{k}\,,
\ee 
the symbol $\eta_{123}$ denotes an odd conformally invariant cross-ratio
\be
\label{oddCrossRatio}
\eta_{123}=\frac{\theta_1 Z_{23}+\theta_2 Z_{31}+\theta_3 Z_{12}+\theta_1\theta_2\theta_3}{
(Z_{12}Z_{13}Z_{32})^{\frac{1}{2}}
}
\ee
and $C_{h_1h_2h_3}$, $\tilde{C}_{h_1h_2h_3}$ are two independent three-point structure constants.

A Verma supermodule associated with a superfield $\Phi_{\Delta_i}$ will be denoted by
\begin{equation} \label{thetorus0-1}
\mathcal{H}_{\Delta_i}=V_{\Delta_i} \otimes \overline{V}_{\Delta_{i}},
\end{equation}
and is spanned by the basis of descendant states
\begin{equation}  \label{thetorus0-2}
\begin{aligned}
& |M, \bar{M}, \Delta_i\rangle=L_{-n_{1}}^{i_{1}} \cdots L_{-n_{l}}^{i_{l}} G_{-r_{1}}^{j_{1}} \cdots  G_{-r_{k}}^{j_{k}} \times \bar{L}_{-\bar{n}_{1}}^{\bar{i}_{1}} \cdots \bar{L}_{-\bar{n}_{l}}^{\bar{i}_{l}} \bar{G}_{- \bar{r}_{1}  }^{\bar{j}_{1}}   \cdots \bar{G}_{- \bar{r}_{k} }^{\bar{j}_{k}}  |\Delta_i\rangle,\\
& |M|=n_{1} i_{1}+\cdots+n_{l}  i_{l}+r_{1}j_1+\cdots+ r_{k} j_{k},\\
& |\bar{M}|=\bar{n}_{1} \bar{i}_{1}+\cdots+\bar{n}_{l}  \bar{i}_{l}+\bar{r}_{1} \bar{j}_{1}+\cdots +\bar{r}_{k} \bar{j}_{k}, \quad |M|,  |\bar{M}| \in \frac{\mathbb{N}^{+}}{2} ,
\end{aligned}
\end{equation}
where 

\begin{equation}  \label{thetorus0-3}
|\Delta_i\rangle=\Phi_{\Delta_i}(0,0)|0\rangle
\end{equation}
is the highest weight state, which satisfies $
 L_0 |\Delta_i\rangle=\Delta_i|\Delta_{i}\rangle$, and it is annihilated by the generators $ L_n, \bar{L}_n$ for $n > 0$. $\mathbb{N}^{+}$ stands for non negative integers. If one focuses only on the holomorphic sector $V_{\Delta_i}$ of (\ref{thetorus0-1}), then the descendant states of the supermodule $V_{\Delta_i}$  can be written as

\begin{equation}  \label{thetorus0-4}
|M, \Delta_i\rangle= L_{-n_{1}}^{i_{1}} \cdots L_{-n_{l}}^{i_{l}} G_{-r_{1}}^{j_{1}} \cdots  G_{-r_{k}}^{j_{k}}|\Delta_i\rangle.
\end{equation}
To compute the global conformal blocks it is sufficient to consider only the $\mathfrak{osp} (1|2)$ subsector of the $N=1$ supersymmetric theory. For this, it will be convenient to write the basis of states of the  $\mathfrak{osp} (1|2)$ supermodule $V_{\Delta_i}^{\mathfrak{osp} }$ just as

\begin{equation}  \label{thetorus0-5}
    |M, \Delta_i\rangle=L_{-1}^m G_{-1 / 2}^k|\Delta_i\rangle, \quad M=(m, k), \quad m \in \mathbb{N}^{+}, \quad k=0,1.
\end{equation}
Clearly, the $\mathfrak{osp} (1|2)$ supermodule factorizes into two $\mathfrak{sl}(2)$ modules, as follows

\begin{equation}  \label{thetorus0-6}
    V_{\Delta_i}^{\mathfrak{osp} }=V_{\Delta_i}^{\mathfrak{s l}(2)} \otimes V_{\Delta_i+\frac{1}{2}}^{\mathfrak{s l}(2)},
\end{equation}
where the $V_{\Delta_i}^{\mathfrak{s l} (2)}$ module is spanned by states $|M, \Delta_i \rangle $ with $k=0$ and the $V_{\Delta_i+\frac{1}{2}}^{\mathfrak{s l} (2)}$ with $k=1$, according to the notation (\ref{thetorus0-5}). These two sectors correspond to the even and odd parts of the $\mathfrak{osp} (1|2)$ supermodule, respectively.
\section{Shadow formalism for supersymmetric case}
\label{SuperShadow}
\subsection{Supersymmetric shadow operator}
In the spirit of the original shadow formalism, we seek the shadow operator in the form of an  integral over coordinate space of the corresponding primary multiplied by some function of coordinates, and the projector 
as
\be
\label{projectorDef}
    \Pi_{h,\bar{h}}=\int d^2 w \int d^2 \xi 
    \mathcal{O}_{h,\bar{h}}(W,\xi)\mid 0 \rangle
    \langle 0 \mid \tilde{\mathcal{O}}_{h,\bar{h}}(W,\xi)
\ee
In order for the nonlocal operator of the form 
\be
\label{SuperAnsatz}
	\tilde{\mathcal{O}}_{h,\bar{h}}(z,\bar{z},\theta,\bar{\theta})=\int d^2 w \int d^2 \xi f(z,\bar{z},\theta,\bar{\theta};w,\bar{w},\xi,\bar{\xi})\mathcal{O}_{h,\bar{h}}(w,\bar{w},\xi,\bar{\xi})
\ee
to be a quasi-primary it must be equal to
\be	
\label{superShadowDef}
	\tilde{\mathcal{O}}_{h,\bar{h}}(z,\xi,\bar{z},\bar{\xi})
	=N_{h}\int d^2 w \int d^2 \theta  \frac{\mathcal{O}_{h,\bar{h}}(w,\theta,\bar{w},\bar{\theta})}{
		(z-w-\theta\xi)^{1-2h}(\bar{z}-\bar{w}-\bar{\theta}\bar{\xi})^{1-2\bar{h}}}
\ee
for some value of the normalization constant $N_{h}$.
The superconformal dimension $(h^{*},\bar{h}^{*})$ of the shadow operator(\ref{superShadowDef}) is related to the dimension of the superprimary field $\mathcal{O}$ as follows
\be
\label{SuperShadowDim}
    h^{*}=\frac{1}{2}-h\,,\quad \bar{h}^{*}=\frac{1}{2}-\bar{h}\,.
\ee
Then the operator (\ref{projectorDef}) has superconformal dimension $(0,0)$.
In the remainder of the paper, for the sake of clarity, we will restrict ourselves to the spinless fields, so that
\be
    h=\bar{h}
\ee
and completely omit the dependence on $\bar{h}$. Nevertheless, with minimal effort our results can be generalized to fields with non-zero spin.

The definition (\ref{superShadowDef}) implies the following
relation between the structure constants
\be
\label{SShadow3ptOdd}
    \tilde{C}^{*}_{hh_1h_2}=2^{4h-2}(2h-1)^2 I_0(h_1+h-h_2,h_2+h-h_1) N_h C_{hh_1h_2}\,,
\ee
\be
\label{SShadow3ptEven}
    C^{*}_{hh_1h_2}=2^{4h}I_0\left(h_1+h-h_2+\frac{1}{2},h_2+h-h_1+\frac{1}{2}\right)N_h\tilde{C}_{hh_1h_2}\,,
\ee
where $C^{*}_{hh_1h_2}$, $\tilde{C}^{*}_{hh_1h_2}$ are three-point constants corresponding to the correlation function involving the shadow operator $\tilde{O}_h$ and two superprimary fields $\Phi_{h_1}$ and  $\Phi_{h_2}$. The function $I_0(h_1,h_2)$ in the last equation is defined as follows
\be
	I_0(h_1,h_2)=
	4^{-h_1-h_2+1}\pi \frac{\Gamma(1-h_1)\Gamma(1-h_2)\Gamma(h_1+h_2-1)}{\Gamma(h_1)\Gamma(h_2)\Gamma(2-h_1-h_2)}\,,
\ee
and admits the following integral representation
\be
\label{IntRep}
    I_0(h_1,h_2)=\int d^2 w |w-1|^{-2h_1}|w+1|^{-2h_2}
\ee
within the domain of convergence of the r.h.s. of  the eq.(\ref{IntRep}). 

\subsection{Identity decomposition}
One can prove that the operator (\ref{projectorDef}) 
is a projector onto irreducible highest weight $\mathfrak{osp}(1|2)$ module by demonstrating that (\ref{DeltaVir}) can be generalized to supersymmetric case as follows:
\be
\label{SuperDelta}
    \langle \tilde{\mathcal{O}}_h(Z_1,\bar{Z}_1)\mathcal{O}_h(Z_2,\bar{Z}_2)\rangle
    =\delta^2(z_1-z_2)\delta^2(\theta_1-\theta_2)\,.
\ee
Indeed, consider the integral representation
\be
\label{superDeltaInt}
    \langle \tilde{\mathcal{O}}_h(Z_1,\bar{Z}_1)\mathcal{O}_h(Z_2,\bar{Z}_2)\rangle
    =
    N_h\int d^2 \xi d^2 w
    |z_1-w-\theta_1\xi|^{-2+4h}
    |w-z_2-\xi\theta_2|^{-4h}\,.
\ee
As it is clearly divergent, we regularize this expression by replacing $h$ with 
\be
    h_{\epsilon}=h-\frac{\epsilon}{2}
\ee
in the first factor of the integrand in the r.h.s. of (\ref{superDeltaInt}):
\be
    \langle \tilde{\mathcal{O}}_{h}(Z_1,\bar{Z}_1)\mathcal{O}_h(Z_2,\bar{Z}_2)\rangle_{\epsilon}
    =
    N_h\int d^2 \xi d^2 w
    |z_1-w-\theta_1\xi|^{-2+4h-2\epsilon}
    |w-z_2-\xi\theta_2|^{-4h}\,.
\ee
 The integral above can be computed explicitly:
\be
\begin{split}
\langle \tilde{\mathcal{O}}_{h}(Z_{1},\bar{Z}_1) &\mathcal{O}_{h}(Z_{2},\bar{Z}_2)\rangle_{\epsilon}=
\left(\frac{|z_{12}|}{2}\right)^{-2+2\epsilon}\\
\times \Big[&
-4h^2\theta_2\bar{\theta}_2 I_0(2h^{*}_{\epsilon},2h+1)
-2hh^{*}_{\epsilon}(\theta_1\bar{\theta}_2+\theta_2\bar{\theta}_1)I_0(2h^{*}_{\epsilon},2h+1)\\
&-2hh^{*}_{\epsilon}(\theta_1\bar{\theta}_2+\theta_2\bar{\theta}_1)I_0(1+2h^{*}_{\epsilon},2h)
-4(h^{*}_{\epsilon})^2\theta_1\bar{\theta}_1 I_0(1+2h^{*}_{\epsilon},2h)\\
&+8hh^{*}_{\epsilon}(\theta_1\bar{\theta}_2+\theta_2\bar{\theta}_1)
I_0(1+2h^{*}_{\epsilon},1+2h)
\Big]\,,
\end{split}
\ee
where we've used shortcut notation
\be
    h_{\epsilon}^{*}=\frac{1}{2}-h_{\epsilon}\,.
\ee
In the limit $\epsilon\rightarrow 0$ the function $I_0$
is proportional to two-dimensional delta function \cite{Dolan2011}, while delta function of Grassmann variables is simply a linear function:
\be
    \delta(\theta_1-\theta_2)=\theta_1-\theta_2\,, \quad
    \int d\theta_1 f(\theta_1)\delta(\theta_1-\theta_2)=f(\theta_2)\,.
\ee
Therefore, if the normalization factor $N_h$ is chosen as
\be
    N_h=-\frac{1}{4\pi^2}\,,
\ee
then, in this limit, the two-point function becomes a product of delta functions:
\be
    \langle \tilde{\mathcal{O}}_{h}(z_1,\theta_1) \mathcal{O}_{h}(z_2,\theta_2)\rangle
    =
    \delta^2(z_1-z_2)\delta^2(\theta_1-\theta_2)\,.
\ee
Thus, the identity operator admits the decomposition into a sum of projectors (\ref{projectorDef})
\be
\label{SuperIdentity}
    I=\sum_{h}\Pi_{h}\,.
\ee
\section{Supersymmetric conformal blocks}
\label{CBExamples}
\subsection{Four-point conformal block on a sphere}
The supersymmetric four-point correlation function
    of spinless primary superfields can be expressed as
\be
\label{4ptExpansion}
\begin{split}
        \langle \prod\limits_{i=1}^{4}\Phi_{h_i}(Z_i,\bar{Z}_i)\rangle=
        &\left|\mathcal{L}_{h_1,h_2,h_3,h_4}(Z_{12},Z_{34},Z_{24},Z_{13})\right|^2 \\
        &\times\sum_{h} G_h(h_1,h_2,h_3,h_4|X,\bar{X},\eta,\bar{\eta},\eta^{\prime},\bar{\eta}^{\prime})\,,
\end{split}
\ee
where the supercoordinates read in components
\be
    Z_i=(z_i,\theta_i)\,,\quad \bar{Z}_i=(\bar{z}_i,\bar{\theta}_i)\,,
\ee
the variables $X$, $\eta$ and $\eta^{\prime}$ are $OSp(1|2)$ invariant cross-ratios
\be
        X=\frac{Z_{34} Z_{21}}{Z_{31} Z_{24}}\,,
\ee
\be
        \eta=\eta_{124}\,,\quad\eta^{\prime}=(1-X)^{\frac{1}{2}}\eta_{123}\,,
\ee
and the symbol $\mathcal{L}_{h_1,h_2,h_3,h_4}$
\be
\label{legfactor4}
        \mathcal{L}_{h_1,h_2,h_3,h_4}=Z_{12}^{-h_1-h_2}Z_{34}^{-h_3-h_4}Z_{24}^{h_1-h_2}Z_{13}^{-h_3+h_4}Z_{14}^{-h_{1}+h_2+h_3-h_4}
\ee
stands for the ``leg factor'' ensuring correct transformation properties w.r.t. global superconformal transformations,
and $G_h$ is a superconformal block.
Restoring the antiholomorphic dependence and taking into account that the whole correlation function must be even, one gets 
the following general form of the conformal block:
\be
\label{4planarAnsatz}
\begin{split}
        G_h(X,\bar{X},\eta,\bar{\eta},\eta^{\prime},\bar{\eta}^{\prime})=&
        g^{(0,0)}_h(X,\bar{X})
        +g^{(1,0)}_h(X,\bar{X})\eta\bar{\eta}+f^{(1,1)}_h(X,\bar{X})\eta\eta^{\prime}+f^{(-1,-1)}_h(X,\bar{X})\bar{\eta}\bar{\eta}^{\prime}\\
        &+f^{(1,-1)}_h(X,\bar{X})\eta\bar{\eta}^{\prime}+f^{(-1,1)}_h(X,\bar{X})\bar{\eta}\eta^{\prime}\\
        &+g^{(0,1)}_h(X,\bar{X})\eta^{\prime}\bar{\eta}^{\prime}
        +g^{(1,1)}_h(X,\bar{X})\eta\eta^{\prime}\bar{\eta}\bar{\eta}^{\prime}\,.
\end{split}
\ee
Inserting identity decomposition (\ref{SuperIdentity})
between $\Phi_{h_2}$ and $\Phi_{h_3}$ we represent the four-point function as a sum of superconformal partial waves
\be
    \langle \prod\limits_{i=1}^{4}\Phi_{h_i}(Z_i,\bar{Z}_i)\rangle=
    \sum_{h}\Psi^{h_1,\dots,h_4}_{h}(Z_1,\dots,Z_4;\bar{Z}_1,\dots,\bar{Z}_4)\,,
\ee
\be
\begin{split}
    \Psi^{h_1,\dots,h_4}_{h}(Z_1,\dots,Z_4;\bar{Z}_1,\dots,\bar{Z}_4)=
    \int d^2 z_0 \int d^2 \theta_0 \,&\SthreeP_{h_1,h_2,h}(Z_1,\bar{Z}_1;Z_2,\bar{Z}_2;Z_0,\bar{Z}_0)\\
    &\times \SthreeP_{h^{*},h_3,h_4}(Z_0,\bar{Z}_0;Z_3,\bar{Z}_3;Z_4,\bar{Z}_4)\,.
\end{split}
\ee
To compute the components $g^{(i,j)}_h(X,\bar{X})$ appearing in the expansion (\ref{4planarAnsatz}) we can set some of the Grassmann coordinates $\theta_i$ and $\bar{\theta}_i$ to zero, reducing functions of $X,\bar{X}$ to functions of $x,\bar{x}$, where
\be
    x=\frac{z_{43} z_{21}}{z_{31}z_{24}}
\ee
and use the superanalyticity (\ref{SuperAnalytic}) of the conformal block to restore dependence on Grassmann variables.
In particular, the function $g^{(0,0)}_h(X,\bar{X})$ can be computed by setting all four holomorphic Grassmann variables, along with their antiholomorphic counterparts, to zero.
At this point of the superspace the invariant cross-ratios evaluate to complex numbers:
\be
    X=x\,, \quad\eta=0\,,\quad\eta^{\prime}=0\,.
\ee
The conformal partial wave then admits the following integral representation
\be
  \begin{split}  \Psi^{h_1,\dots,h_4}_h(z_1,\dots,z_4;\bar{z}_1,\dots,\bar{z}_4)=
    &\int d^2 z_0 
    \frac{\tilde{C}_{12h}C_{h34}^{*}|z_{12}|^{2h-2h_1-2h_2+1}|z_{34}|^{1-2h-2h_3-2h_4}}{
    |z_{10}|^{1+2p_1}|z_{20}|^{1+2p_2}
    |z_{03}|^{2p_3}|z_{04}|^{2p_4}}\\
     &+\int d^2 z_0 
    \frac{C_{12h}\tilde{C}_{h34}^{*}|z_{12}|^{2h-2h_1-2h_2)} |z_{34}|^{2-2h-2h_3-2h_4}}{|z_{10}|^{2p_1}|z_{20}|^{2p_2}|z_{03}|^{1+2p_3}|z_{04}|^{1+2p_4}}
\end{split}
\ee
with the powers in the denominators given by
\be
	p_1=h+h_{12}\,,\quad p_2=h-h_{12}\,,\quad p_3=\frac{1}{2}-h+h_{34}\,,\quad p_4=\frac{1}{2}-h-h_{34}\,,
\ee
and the shortcut notation $h_{ij}$ is used for the difference of conformal dimensions:
\be
    h_{ij}=h_{i}-h_{j}\,.
\ee
The conformal partial wave is known to contain contributions from the global conformal block $G_h$ and the so-called shadow global conformal block $G_{h^{*}}$ \cite{Osborn2012}, and this holds true for the superconformal partial waves as well, as will be demonstrated by the calculations below.
Performing a linear fractional transformation mapping the points $z_1,z_2,z_3,z_4$ to $\infty,1,x,0$ respectively
and stripping off the leg factor we obtain the following expression for the function $g^{(0,0)}_h$
\be
\label{VVVVan}
\begin{split}
    g^{(0,0)}_h(X,\bar{X})+g^{(0,0)}_{h^{*}}(X,\bar{X})=&\tilde{C}_{hh_1h_2}C^{*}_{hh_3h_4}|X|
    \hgInt\left(h+\frac{1}{2},h_{12},h_{34}\Big|X,\bar{X}\right)\\
    &+C_{hh_1h_2}\tilde{C}^{*}_{hh_3h_4}|X|
    \hgInt\left(h,h_{12},h_{34}|X,\bar{X}\right)\;,
\end{split}
\ee
where
\be
\begin{split}
    \hgInt(h,h_{12},h_{34}|X,\bar{X})=&
    \hgCoeff(1-h,h_{12})\left|X^{\frac{1}{2}-h}{}_2F_1(1-h+h_{34},1-h-h_{12},2-2h|X)\right|^2
    \\
    &+\hgCoeff(h,h_{34})\left|X^{h-\frac{1}{2}}{}_2F_1(h+h_{34},h-h_{12},2h|X)\right|^2.
\end{split}
\ee
Here, ${}_2F_1$ stands for the hypergeometric function, and the coefficients $\hgCoeff$ are given by
\be
    \hgCoeff(h,h^{\prime})=
    \pi \frac{\Gamma(1-2h)\Gamma(h+h^{\prime})\Gamma(h-h^{\prime})}{\Gamma(2h)\Gamma(1-h-h^{\prime})\Gamma(1-h+h^{\prime})}\;.
\ee
Taking into account relations (\ref{SShadow3ptOdd},\ref{SShadow3ptEven}) between the structure constants one can verify that
the function $\hgInt$ satisfies the following identity
\be
    C^{*}_{hh_1h_2}\tilde{C}_{hh_3h_4}\hgInt\left(\frac{1}{2}-h,h_{12},h_{34}\Big|X,\bar{X}\right)=
    \tilde{C}_{hh_1h_2}C^{*}_{hh_3h_4}\hgInt\left(h+\frac{1}{2},h_{12},h_{34}\Big|X,\bar{X}\right)\;.
\ee
Applying this identity to the r.h.s. of eq. (\ref{VVVVan}) we see that the conformal partial wave is invariant under interchange of conformal dimension of the intermediate primary field  $h$ and its shadow dual  $h^{*}$.
 Therefore, we are justified in interpreting this result as a sum of contributions from the global conformal block and the shadow global conformal block. The former can be extracted from the whole conformal partial wave by selecting terms with correct asymptotic behaviour. Explicitly, the function $g^{(0,0)}_h$ can be expressed as follows:
\be
\begin{split}
    g^{(0,0)}_h(X,\bar{X})=& \tilde{C}_{012}\tilde{C}_{034}\left|G_{0,0}^{(o)}(h,h_{12},h_{34}|X)\right|^2
    +C_{012}C_{034}
     \left|G_{0,0}^{(e)}(h,h_{12},h_{34}|X)\right|^2\,,
\end{split}
\ee
where
\be
\label{4ptVVVVeven}
    G_{0,0}^{(e)}(h,h_{12},h_{34}|X)=
    X^h\,{}_2F_1(h_{34}+h,-h_{12}+h,2h|X)\,,
\ee
\be
\label{4ptVVVVodd}
    G_{0,0}^{(o)}(h,h_{12},h_{34}|X)=
    \frac{1}{2h}G^{(e)}_{0,0}\left(h+\frac{1}{2},h_{12},h_{34}\Big|X\right)\,.
\ee
The expressions (\ref{4ptVVVVeven}),(\ref{4ptVVVVodd}) for the components of the four-point superconformal block are (up to a choice of order of points $z_1,z_2,z_3,z_4$) in agreement with earlier results \cite{Belavin2007,Belavin2008,Hikida2018}.

\section{Torus superconformal blocks} \label{sec:torussb}
In the following sections, we study superconformal field theory on a two-dimensional torus. We will focus on the one and two-point torus superconformal blocks. Our goal is to describe the torus superconformal blocks using shadow formalism.

Similarly to the spherical case, the correlation functions on torus can be decomposed into a sum over intermediate primary superfields, but even the one-point function on a torus is already nontrivial (i.e., involves a sum over all superprimary fields in the theory).
The one-point and two-point torus correlation functions of primary superfields $\Phi_{h_1}(Z_{1}, \bar{Z}_{1}), \Phi_{h_2}(Z_{2}, \bar{Z}_{2})$ can be written as
\begin{equation}\label{thetorus1}
\begin{aligned}
& \langle\Phi_{h_1}(Z_{1}, \bar{Z}_{1})
\rangle_{\tau}=\sum_{\Delta_1} \operatorname{str}_{ \mathcal{H}_{\Delta_1}  }\left[q^{L_0} \bar{q}^{\bar{L}_{0}} \Phi_{h_1}(Z_{1}, \bar{Z}_{1})\right] ,\\
\end{aligned}
\end{equation}
\begin{equation} \label{thetorus2}
    \begin{aligned}
        \langle \Phi_{h_1}(Z_{1}, \bar{Z}_{1}) \Phi_{h_2}(Z_{2}, \bar{Z}_{2})\rangle_{\tau}=\sum_{\Delta_1} 
 \operatorname{str}_{ \mathcal{H}_{\Delta_1}  } \left[q^{L_0} \bar{q}^{\bar{L}_{0}}  \Phi_{h_1}(Z_{1}, \bar{Z}_{1}) \Phi_{h_2}(Z_{2}, \bar{Z}_{2})\right],
    \end{aligned}
\end{equation}
where the $ \operatorname{str}_{ \mathcal{H}_{\Delta_1}  } $ stands for the supertrace taken over the supermodule $\mathcal{H}_{\Delta_1}  $. Considering (\ref{superfieldDecomp}), one can decompose the torus correlation functions (\ref{thetorus1},\ref{thetorus2}) into different components obtained from the superfields. By parity arguments in Grassmann variables, only terms with an even number of Grassmann variables contribute\footnote{This argument can be verified by noting that the parity, w.r.t the parity operator $(-1)^F\otimes (-1)^{\bar{F}}$, of the matrix elements representing the structure constants $C_{\Delta_ih_i \Delta_i}, \tilde{C}_{\Delta_ih_i \Delta_i}$ is always even. Notice also that the same behaviour occurs in the sphere two-point function (\ref{2ptsupersymmetric}) and four-point function (\ref{4planarAnsatz}), where only even terms in Grassmann variables contribute.}. Thus, the one-point function (\ref{thetorus1}) can be written as
\begin{equation} \label{thetorus3}
\langle\Phi_{h_1}(Z_{1}, \bar{Z}_{1})\rangle_{\tau}=\langle\phi_{h_1}(z_{1}, \bar{z}_{1})\rangle_{\tau}+\theta_{1} \bar{\theta}_{1}\langle\tilde{\phi}_{h_1}(z_{1}, \bar{z}_{1})\rangle_{\tau}.
\end{equation}
For the two-point function (\ref{thetorus2}), the decomposition reads
\begin{equation} \label{thetorus3-1}
\begin{aligned}
&\langle\Phi_{h_1}(Z_{1}, \bar{Z}_{1}) \Phi_{h_2}(Z_{2}, \bar{Z}_{2})\rangle_{\tau}=\\&
=\langle\phi_{h_1}(z_{1}, \bar{z}_{1}) \phi_{h_2}(z_{2}, \bar{z}_{2})\rangle_{\tau}+\theta_{1} \theta_{2}\langle\psi_{h_1}(z_{1}, \bar{z}_{1}) \psi_{h_2}(z_{2}, \bar{z}_{2})\rangle_{\tau}+\theta_{1} \bar{\theta}_{1}\langle\tilde{\phi}_{h_1}(z_{1}, \bar{z}_{1}) \phi_{h_2}(z_{2}, \bar{z}_{2})\rangle_{\tau}\\&+ \bar{\theta}_{1} \bar{\theta}_{2}\langle \bar{\psi}_{h_1}(z_{1}, \bar{z}_{1}) \bar{\psi}_{h_2}(z_{2}, \bar{z}_{2})\rangle_{\tau}
+\theta_{2} \bar{\theta}_{2}\langle \phi_{h_1}(z_{1}, \bar{z}_{1}) \tilde{\phi}_{h_2}(z_{2}, \bar{z}_{2})\rangle_{\tau}+\theta_{1} \bar{\theta}_{2}\langle\psi_{h_1}(z_{1}, \bar{z}_{1}) \bar{\psi}_{h_2}(z_{2}, \bar{z}_{2})\rangle_{\tau} \\&+ \bar{\theta}_{1} \theta_{2}\langle\bar{\psi}_{h_1}(z_{1}, \bar{z}_{1}) \psi _{h_2}(z_{2}, \bar{z}_{2})\rangle_{\tau}+\theta_{1} \bar{\theta}_{1} \theta_{2} \bar{\theta}_{2}\langle \tilde{\phi}_{h_1}(z_{1}, \bar{z}_{1}) \tilde{\phi}_{h_2}(z_{2}, \bar{z}_{2}) \rangle.
\end{aligned}
\end{equation}
In turn, each of the terms on the r.h.s of (\ref{thetorus3},\ref{thetorus3-1}) can be decomposed into torus superconformal blocks, as we will discuss.
\subsection{One-point superconformal blocks}
For the one-point function (\ref{thetorus3}), the decomposition into superconformal blocks reads

\begin{equation} \label{thetorus4}
\begin{aligned}
& \langle\phi_{h_1}(z_{1}, \bar{z}_{1})\rangle_{\tau}=\sum_{\Delta_1} C_{\Delta_1 h_1   \Delta_1}|B_{0}(h_1, \Delta_1 \mid q)|^{2}, \\
& \langle\tilde{\phi}_{\Delta_1}(z_{1}, \bar{z}_{1})\rangle_{\tau}=\sum_{\Delta_1} \tilde{C}_{ \Delta_1 h_1 \Delta_1 }|B_{1}(h_1, \Delta_1 \mid q)|^{2},
\end{aligned}
\end{equation}
where $B_{0}$ and $ B_{1}$ are the holomorphic one-point lower and upper superconformal blocks

\begin{equation} \label{thetorus5}
\begin{aligned}
& B_{0}(h_1, \Delta_1 \mid q)=\frac{1}{\langle\Delta_1|\phi_{h_1}| \Delta_1\rangle}\operatorname{str}_ {\Delta_1 }\left[q^{L_{0}} \phi_{h_1}\right], \\
& B_{1}(h_1, \Delta_1 \mid q)=\frac{1}{\langle\Delta_1|\psi_{h_1}| \Delta_1\rangle}\operatorname{str}_{\Delta_1 }\left[q^{L_0} \psi_{h_1}\right].
\end{aligned}
\end{equation}
Here the graded supertrace $\operatorname{str}_{\Delta_1}$ is evaluated over the supermodule $V_{\Delta_1}$, and we have used the fact that one can work with $\psi_{h_1}$ to describe purely holomorphic contribution from $\tilde{\phi}_{h_1}$. In the $\mathfrak{osp}(1|2)$ sector, one can compute closed-form expressions for the trace (see \cite{Alkalaev2018}). Thus, one can write explicitly (\ref{thetorus5}) in this sector as

\begin{equation}\label{thetorus5-1}
\begin{aligned}
& B_{0}(h_1, \Delta_1 \mid q)=\frac{1}{\langle\Delta_1|\phi_{h_1}| \Delta_1\rangle} \sum_{n=\frac{\mathbb{N}^{+}}{2}} \sum_{|N|=|M|=n}(-1)^{2 n} \mathcal{B}_{\Delta_1}^{N | M}\langle\Delta_1, N| q^{L_0}\phi_{h_1}| M, \Delta_1\rangle,\\
& B_{1}(h_1, \Delta_1 \mid q)=\frac{1}{\langle\Delta_1|\psi_{h_1}| \Delta_1\rangle} \sum_{n=\frac{\mathbb{N}^{+}}{2}} \sum_{|N|=|M|=n} \mathcal{B}_{\Delta_1}^{N \mid M}\langle\Delta_1, N| q^{L_0}\psi_{h_1}| M, \Delta_1\rangle,
\end{aligned}
\end{equation}
where the sum is over the states (\ref{thetorus0-5}), and $\mathcal{B}^{N|M}_{\Delta_i}$ is matrix element ($N,M$) of the inverse of the Gram matrix. Closed-form expressions for (\ref{thetorus5-1}) are given in terms in linear combinations of $\mathfrak{sl}(2)$ one-point torus conformal blocks, namely

\begin{equation}\label{thetorus5-2}
\begin{aligned}
& B_{0}(h_1, \Delta_1 \mid q)=\mathcal{F}_{\Delta_1}^{h_1}(q)-\frac{(2 \Delta_1-h_1)}{2 \Delta_1} \mathcal{F}_{\Delta_1+\frac{1}{2}}^{h_1}(q) ,\\
& B_{1}(h_1, \Delta_1 \mid q)=\quad \mathcal{F}_{\Delta_1}^{h_1+\frac{1}{2}}(q)-\frac{(2 \Delta_1+h_1-\frac{1}{2})}{2 \Delta_1} \mathcal{F}_{\Delta_1+\frac{1}{2}}^{h_1+\frac{1}{2}}(q) ,
\end{aligned}
\end{equation}
where $\mathcal{F}_{\Delta_1}^{h_1}(q)$ is the $\mathfrak{sl}(2)$ one-point torus conformal block
\begin{equation} \label{thetorus5-3}
    \mathcal{F}_{\Delta_1}^{h_1}(q)=\frac{q^{\Delta_1}}{(1-q)^{1-h_1}}{ }_{2} F_{1}(h_1, h_1+2 \Delta_1-1,2 \Delta_1 \mid q).
\end{equation}
We notice that the linear combinations (\ref{thetorus5-2}) can be obtained by splitting the sum over $n$ in (\ref{thetorus5-1}) into the even and odd parts according to (\ref{thetorus0-6}). Each term obtained after the splitting can be written in terms of (\ref{thetorus5-3}), and using the relations between matrix elements (\ref{thetorus18-0}, \ref{thetorus18}, \ref{thetorus19}), one can obtain precisely (\ref{thetorus5-2}). The same idea can be applied to higher-point superconformal blocks.

\subsection{Two-point superconformal blocks}

Higher-point correlation functions can be decomposed into superconformal blocks in different channels. In this work, we are interested in the necklace channel decompositions. For this, one inserts the following resolution of identity  between the primary fields
\begin{equation} \label{thetorus7}
\mathbb{I}=\sum_{\Delta_2}  \sum_{n,m \in \frac{\mathbb{N}^+}{2}  }^{\infty} \sum_{ |M|=|N|=n} \sum_{|\bar{M}|=|\bar{N}|=m} \quad|N, \bar{N}, \Delta_2\rangle \mathcal{B}^{N | M}_{\Delta_2}\bar{\mathcal{B}}^{ \bar{N} \mid \bar{M}}_{\Delta_2}\langle\Delta_2, M, \bar{M}|,
\end{equation}
into the terms of r.h.s of (\ref{thetorus3-1}).
Let us explain this construction for the first and second terms of (\ref{thetorus3-1}). The discussion for the other terms follows the same idea. For the purely bosonic contribution, the decomposition into conformal blocks can be written as

\begin{equation} \label{thetorus8}
\begin{aligned}
& \langle\phi_{h_1}(z_{1}, \bar{z}_{1}) \phi_{h_2}(z_{2}, \bar{z}_{2})\rangle_{\tau} \\
& =\sum_{\Delta_1}\operatorname{str}_{\mathcal{ H}_{\Delta_1} }\left[q^{L_{0}} \bar{q}^{\bar{L}_{0}} \phi_{h_1}(z_{1}, \bar{z}_{1})\mathbb{I} \phi_{h_2}(z_{2}, \bar{z}_{2})\right] \\
& =\sum_{\Delta_1, \Delta_2} C_{\Delta_1 h_1 \Delta_2} C_{\Delta_2 h_2 \Delta_1}|B_{00}^{(1)}(q_{1}, z_{1}, z_{2})|^{2}+\tilde{C}_{\Delta_1 h_1 \Delta_2} \tilde{C}_{\Delta_2 h_2 \Delta_1}|B_{00}^{(2)}(q_{1}, z_{1}, z_{2})|^{2},
\end{aligned}
\end{equation}
where $B_{00}^{(1)}$, and $B_{00}^{(2)}$ are the holomorphic two-point superconformal blocks, which provide contributions from odd and even parts of the trace $\operatorname{str}_{\mathcal{ H}_{\Delta_1} }$ and the operator (\ref{thetorus7}). These two conformal blocks are given by
\begin{equation} \label{thetorus9}
\begin{aligned}
& B_{00}^{(1)}(q, z_{1}, z_{2})=\frac{  z_1^{\Delta_1-\Delta_2-h_1} z_2^{-\Delta_1+\Delta_2-h_2} }{\langle\Delta_1|\phi_{h_1}(z_1)| \Delta_2\rangle\langle\Delta_2|\phi_{h_2}(z_2)| \Delta_1\rangle} \\& \times \sum_{\substack{n, m=0 \\ n+m=\mathbb{N}}} \sum_{\substack{|M|=|N|=n \\|S|=|T|=m}}  (-1)^{2n}
 \mathcal{B}_{\Delta_1}^{M \mid N}\langle\Delta_1 ,M|q^{L_0} \phi_{h_1}(z_1)| S, \Delta_2\rangle  \mathcal{B}_{\Delta_2}^{S \mid T}\langle\Delta_2, T|\phi_{h_2}(z_2)| N, \Delta_1\rangle ,
\end{aligned}
\end{equation}

\begin{equation} \label{thetorus10}
\begin{aligned}
& B_{00}^{(2)}(q, z_{1}, z_{2})=\frac{z_1^{\Delta_1-\Delta_2-h_1-\frac{1}{2}} z_2^{-\Delta_1+\Delta_2-h_2+\frac{1}{2}}} {\langle\Delta_1|\phi_{h_1}(z_1)| \Delta_2 + \frac{1}{2}\rangle\langle\Delta_2+ \frac{1}{2}|\phi_{h_2} (z_2)| \Delta_1\rangle}  \\& \times \sum_{\substack{n, m=0 \\ 2(n+m)=\text{odd} } } \sum_{\substack{|M|=|N|=n \\|S|=|T|=m}}  (-1)^{2n}
 \mathcal{B}_{\Delta_1}^{M \mid N}\langle\Delta_1 ,M|q^{L_0} \phi_{h_1}(z_1)| S, \Delta_2\rangle   \mathcal{B}_{\Delta_2}^{S \mid T}\langle\Delta_2, T|\phi_{h_2}(z_2)| N, \Delta_1\rangle ,
\end{aligned}
\end{equation}
In (\ref{thetorus9}, \ref{thetorus10}) and below, given that we mostly emphasize the holomorphic dependence, we omit the antiholomorphic coordinate of the field, and we also use the notion that the holomorphic coordinate dependence of the matrix element $\langle \Delta_1| \phi_{h_1} (z_i)| \Delta_2 \rangle$ is given by $\langle \Delta_1| \phi_{h_1} (z_i)| \Delta_2 \rangle = \langle \Delta_1| \phi_{h_1} (1)| \Delta_2 \rangle  z_i^{\Delta_1-h_1-\Delta_2}$. Notice that in (\ref{thetorus9}), the condition $ n+m=\mathbb{N} $ indicates that sum over $ n$ and $m $ is performed such that  $2n$ and $2 m$  have the same parity, while in (\ref{thetorus10}) the condition  $2(n+m)= odd$ indicates that the parity of  $2 m $ and $2 n$ is different. 

For the second term of (\ref{thetorus3-1}), we have
\begin{equation}  \label{thetorus11}
\begin{aligned}
& \langle\psi_{h_1}(z_{1}, \bar{z}_{1}) \psi_{h_2}(z_{2}, \bar{z}_{2})\rangle_{\tau}=\sum_{\Delta_1}\operatorname{str}_{\mathcal{H}_{\Delta_1}}\left[q^{L_{0}} \bar{q}^{\bar{L}_{0}} \psi_{h_1}(z_{1}, \bar{z}_{1}) \mathbb{I}  \psi_{h_2}(z_{2}, \bar{z}_{2})\right] =\\
& =\sum_{\Delta_1, \Delta_2} \tilde{C}_{\Delta_1 h_1 \Delta_2} \tilde{C}_{\Delta_2 h_2\Delta_1}B_{\theta_{1} \theta_{2}}^{(2)}(q, z_{1}, z_{2})\bar{ B}_{00}^{(2)}(\bar{q}, \bar{z}_{1},\bar{z}_{2})\\&+C_{\Delta_1 h_1 \Delta_2 } C_{\Delta_2 h_2  \Delta_1} B_{\theta_{1} \theta_{2}}^{(1)}(q, z_{1}, z_{2})\bar{ B}_{00}^{(1)}(\bar{q}, \bar{z}_{1},\bar{z}_{2}),
\end{aligned}
\end{equation}
where the two holomorphic superconformal blocks $
B_{\theta_{1} \theta_{2}}^{(2)}(q, z_{1}, z_{2}), B_{\theta_{1} \theta_{2}}^{(1)}(q, z_{1}, z_{2})
$
similarly to (\ref{thetorus9}, \ref{thetorus9}), are given by

\begin{equation}  \label{thetorus12}
\begin{aligned}
 B_{\theta_{1} \theta_{2}}^{(1)}(q, z_{1}, z_{2})& = \frac{   z_1^{\Delta_1-\Delta_2-h_1-1} z_2^{-\Delta_1+\Delta_2-h_2}}{\langle\Delta_1|\psi_{h_1} (z_1)| \Delta_2+\frac{1}{2}\rangle\langle\Delta_2+\frac{1}{2}|\psi_{h_2}(z_2)| \Delta_1\rangle}  \\& \times \sum_{\substack{n, m=0 \\
2(n+m)= od d}} \sum_{\substack{|M|=|N|=n\\ |S|=|T|=m}} (-1)^{2n}\mathcal{B}_{\Delta_1}^{M \mid N}\langle\Delta_1, M|q^{L_{0}} \psi_{h_1}(z_{1})| S, \Delta_2\rangle   \\&  \times \mathcal{B}_{\Delta_2}^{S \mid T}\langle\Delta_2, T|\psi_{h_2}(z_{2})| N, \Delta_1\rangle,
\end{aligned}
\end{equation}

\begin{equation}  \label{thetorus13}
\begin{aligned}
& B_{\theta_{1} \theta_{2}}^{(2)}(q, z_{1}, z_{2})=  \frac{ z_1^{\Delta_1-\Delta_2-h_1-\frac{1}{2}} z_2^{-\Delta_1+\Delta_2-h_2-\frac{1}{2}}}{\langle\Delta_1|\psi_{h_1} (z_1)| \Delta_2\rangle\langle\Delta_2|\psi_{h_2}(z_2)| \Delta_1\rangle}  \\ & \times \sum_{\substack{n, m=0 \\
n+m= \mathbb{N}}} \sum_{\substack{|M|=|N|=n \\
|S|=|T|=m}} (-1)^{2n}\mathcal{B}_{\Delta_1}^{M \mid N}\langle\Delta_1, M|q^{L_{0}} \psi_{h_1}(z_{1})| S, \Delta_2\rangle   \mathcal{B}_{\Delta_2}^{S \mid T}\langle\Delta_2, T|\psi_{h_2}(z_{2})| N, \Delta_1\rangle.
\end{aligned}
\end{equation}
By concentrating on the $\mathfrak{osp}(1|2)$ submodule, we can utilize various methods to derive closed-form expressions for (\ref{thetorus9}, \ref{thetorus10}, \ref{thetorus11}, \ref{thetorus12}). In the following section, we will employ the shadow formalism to compute them. This technique proved efficient for calculating global higher-point torus conformal blocks \cite{Alkalaev2023}. Since no known expressions exist, we will ensure that the expressions derived using the shadow formalism precisely correspond to the conformal blocks under consideration. On the one hand, we can derive differential equations for superconformal blocks from the $\mathfrak{osp}(1|2)$ Casimir operator \cite{Kraus:2017ezw, Alkalaev2018, Alkalaev2022}, which these superconformal blocks must satisfy. We will analyze this in section~\ref{sec:casimirini}.
 
On the other hand, one can obtain closed-form expressions for the global $\mathfrak{osp}(1|2)$ two-point superconformal blocks by splitting the sum over the descendant states in (\ref{thetorus9}, \ref{thetorus10}, \ref{thetorus11}, \ref{thetorus12}) into $\mathfrak{sl}(2)$ modules, and then express the obtained terms using $\mathfrak{sl}(2)$ two-point torus conformal blocks. Let us explain this simple rationale for the purely bosonic part (\ref{thetorus9}). The sums of (\ref{thetorus9}) can be regrouped in even and odd parts as follows

\begin{equation}  \label{thetorus14}
\begin{aligned}
&  B_{00}^{(1)}(q, z_{1}, z_{2})= \frac{z_1^{\Delta_1-\Delta_2-h_1} z_2^{-\Delta_1+\Delta_2-h_2} }{\langle\Delta_1|\phi_{h_1}(z_1)| \Delta_2\rangle\langle\Delta_2|\phi_{h_2}(z_2)| \Delta_1\rangle}  \\
& \quad \times  \sum_{n, m=\mathbb{N}^{+}} \sum_{\substack{|M|=|N|=n \\ |S|=|T|=m}} \left( \underbrace{\mathcal{B}_{\Delta_ 1}^{M |N}\langle\Delta_1, M|q^{L_0} \phi_{h_1}(z_1)| S, \Delta_2\rangle \mathcal{B}_{\Delta_ 2}^{S | T}\langle\Delta_2, T|\phi_{h_2}(z_2)| N, \Delta_1\rangle}_{(1)} \right.  \\
& \quad - \left.  \underbrace{\mathcal{B}_{\Delta_1+\frac{1}{2}}^{M \mid N}\langle\Delta_1+\frac{1}{2}, M|q^{L_0} \phi_{h_1}(z_1)| S, \Delta_2+\frac{1}{2}\rangle \mathcal{B}_{\Delta_2+\frac{1}{2}}^{S \mid T}\langle \Delta_2+\frac{1}{2}, T|\phi_{h_2}(z_2)| N, \Delta_1+\frac{1}{2}\rangle}_{(2)} \right) ,
\end{aligned}
\end{equation}
where in the second term, we used the notation
\begin{equation}  \label{thetorus15}
|N, \Delta_i+\frac{1}{2}\rangle=(L_{-1})^{|N|} G_{-1 / 2}|\Delta_i\rangle.
\end{equation}
One can check that  first term of (\ref{thetorus14}) is proportional to the $\mathfrak{sl}(2)$ two-point torus conformal block $\mathcal{F}_{\Delta_1, \Delta_2}^{h_1, h_2}(q, z_{1}, z_{2})$ \cite{Alkalaev2017b} given by (\ref{torusshadow4}) while the second term is proportional to $\mathcal{F}_{\Delta_1+\frac{1}{2}, \Delta_2+\frac{1}{2}}^{h_1, h_2}(q, z_{1}, z_{2})$. By using the following relations 

\begin{equation}  \label{thetorus18-0}
 \langle \Delta_i+\frac{1}{2} | \Delta_i+\frac{1}{2}\rangle=2 \Delta_i\langle\Delta_i | \Delta_i\rangle=2 \Delta_i,
\end{equation}

\begin{equation}  \label{thetorus18}
\begin{aligned}
& \langle\Delta_1+\frac{1}{2}|\phi_{h_1}(1)| \Delta_2+\frac{1}{2}\rangle=(\Delta_1+\Delta_2-h_1)\langle\Delta_1|\phi_{h_1}(1)| \Delta_2\rangle,
\end{aligned}
\end{equation}
one gets that (\ref{thetorus14}) is given by the linear combination (\ref{torusshadow13-1}). For the other superconformal blocks, e.g., (\ref{thetorus12}, \ref{thetorus13}), one can repeat the same rationale and obtain similar expressions, which will be detailed below. For  (\ref{thetorus12}, \ref{thetorus13}) one requires the relations
\begin{equation} \label{thetorus19}
\begin{aligned}
&     \langle \Delta_1|  \psi_{h_1}(1)| \Delta_2+  \frac{1}{2}\rangle =    (\Delta_1-h_1-\Delta_2) \langle \Delta_1|  \phi_{h_1}(1)| \Delta_2\rangle,\\ &
 \langle \Delta_1+  \frac{1}{2}|  \psi_{h_1}(1)| \Delta_2\rangle =    (\Delta_1+h_1-\Delta_2) \langle \Delta_1|  \phi_{h_1}(1)| \Delta_2\rangle,\\&
 \langle \Delta_1+  \frac{1}{2}|  \psi_{h_1}(1)| \Delta_2+  \frac{1}{2}\rangle =    -(\Delta_1+h_1+\Delta_2-  \frac{1}{2}) \langle \Delta_1|  \psi_{h_1}(1)| \Delta_2\rangle.
\end{aligned}
\end{equation}

\section{Torus superconformal blocks via shadow formalism} \label{sec:torusbsh1}

In this section, we apply shadow formalism to compute global one- and two-point torus superconformal blocks. The generalization of the shadow formalism to the $\mathfrak{osp}(1|2)$ case follows a similar approach to that of the $\mathfrak{sl}(2)$ case. 
\subsection{ $\mathfrak{osp}(1|2)$ torus shadow formalism}

A straightforward generalization of (\ref{torusshadow1-1}, \ref{torusshadow1}) to the $\mathfrak{osp}(1|2)$ case involves replacing the three-point function $v$ with the supersymmetric three-point function (\ref{3ptsupersymmetric}). Thus, for the supersymmetric case, in analogy with (\ref{torusshadow1-1}, \ref{torusshadow1}), one can define the one- and two-point torus superconformal partial waves as follows

\begin{equation}\label{torusshadow6}
W_{\Delta_{1}}^{h_{1}}\left(q, \bar{q} , Z_1, \bar{Z}_1\right)=q^{\Delta_{1}} \bar{q}^{\Delta_{1}} \int d^{2} w_{1} d^{2} \xi_1 \mathcal{V}_{\Delta_{1}^{*}, h_{1}, \Delta_{1}}\left(W_1, \bar{W}_1 ; Z_{1}, \bar{Z}_{1} ; q \cdot W_1, \bar{q} \cdot \bar{W}_1\right),
\end{equation}

\begin{equation} \label{torusshadow7}
\begin{aligned}
&W_{\Delta_{1}, \Delta_{2}}^{h_{1}, h_{2}}\left(q, \bar{q}, Z_{1}, \bar{Z}_1, Z_{2} , \bar{Z}_2
 \right)=\\&  \begin{split}  q^{\Delta_{1}} \bar{q}^{\Delta_{1}} \int d^{2} w_{1} d^2w_{2}   d^{2} \xi_1 d^{2} \xi_2 & \mathcal{V}_{\Delta_{1}^{*}, h_{1}, \Delta_{2}}\left(W_{1}, \bar{W}_{1} ; Z_{1}, \bar{Z}_{1} ; W_{2}, \bar{W}_{2}\right)  \\& \times \mathcal{V}_{\Delta_{2}^{*}, h_{2}, \Delta_{1}}\left(W_{2}, \bar{W}_{2} ; Z_{2}, \bar{Z}_{2} ; q \cdot W_{1}, \bar{q} \cdot \bar{W}_1\right). \end{split}
 \end{aligned}
\end{equation}
For supersymmetric shadow formalism, we find that the conformal dimension $\Delta_i^*$ of the shadow field is given by the relation (\ref{SuperShadowDim}), and the product $q \cdot W_i$  is defined as follows

\begin{equation} \label{torusshadow8}
\begin{aligned}
& q \cdot W_{i}=(q  w_{i}, \sqrt{q} \xi _i) .
\end{aligned}
\end{equation}
We will see in the discussion below that the definition (\ref{torusshadow8})\footnote{A similar relation to (\ref{torusshadow8})  was found in \cite{Belavin2024} in the discussion of the shadow formalism for $\mathcal{W}_3$ CFT.} is relevant for computing the superconformal blocks.  Since we are interested in the holomorphic superconformal blocks, we will focus only on the holomorphic parts of (\ref{torusshadow6}, \ref{torusshadow7}) and apply the same logic discussed for the global $\mathfrak{sl}(2)$ conformal blocks.

\subsection{One-point torus superconformal blocks}

For the one-point torus superconformal block, we first expand the integrand of (\ref{torusshadow6}) in Grassmann variables and then take the integral over the two variables $\xi_1$ and $\bar{\xi}_1$. This results in expressing (\ref{torusshadow6}) as

\begin{equation} \label{torusshadow9}
W_{\Delta_{1}}^{h_{1}}\left(q,\bar{q}, Z_{1} , \bar{Z}_1\right)=q^{\Delta_{1}} \bar{q}^{\Delta_{1}} \int d^{2} w_{1}\left( \tilde{C}^*_{\Delta_1 h_1 \Delta_1}|b_{0}|^2-\theta_{1} \bar{\theta}_{2} C^*_{\Delta_1 h_1 \Delta_1}| b_{1}|^2\right),
\end{equation}
where $b_1, b_2$ are given by

\begin{equation}\label{torusshadow10}
\begin{aligned}
& b_0=\left(v_{1-\Delta_1, h_1, \Delta_1 }\left(w_1, z_{1}, q w_1\right)+\sqrt{q} v_{\frac{1}{2}-\Delta_1, h_1, \Delta_1+\frac{1}{2}}\left(w_1, z_{1}, q w_1\right)\right), \\&
b_1= \big(( -2 \Delta_1+h_1 +1/2) v_{1-\Delta_1, h_1+\frac{1}{2}, \Delta_1} (w_1, z_1, q w_1) +\\&+(2 \Delta_1+h_1-1/2)\sqrt{q} v_{\frac{1}{2}-\Delta_1, h_1+\frac{1}{2}, \Delta_1+\frac{1}{2}} (w_1, z_1, q w_1) \big).
\end{aligned}
\end{equation}
Integrating (\ref{torusshadow10}) in the same way as discussed for the $\mathfrak{sl}(2)$ case, we obtain that the one-point lower and upper superconformal blocks (\ref{thetorus5-1}) are computed by 

\begin{equation}\label{torusshadow11}
\begin{gathered}
B_{0}\left(h_{1}, \Delta_{1}\mid q\right)=\frac{1}{c_{1}(h_1, \Delta_1)} q^{\Delta_{1}} \int_{0}^{z_{1}} b_0 d w_{1} ,\\
B_{1}\left(h_{1}, \Delta_{1}\mid q\right)=\frac{1}{c_1(h_1+\frac{1}{2}, \Delta_1) ( -2 \Delta_1+h_1 +1/2) }  q^{\Delta_{1}} \int_{0}^{z_{1}} b_1d w_{1}.
\end{gathered}
\end{equation}

\subsection{Two-point torus superconformal blocks}

We repeat the procedure applied to the one-point superconformal blocks to find the two-point torus superconformal blocks. We first expand the integrand of (\ref{torusshadow7}) in the Grassmann variables and then take the integral over $\xi_1, \bar{\xi}_1$ and $\xi_2, \bar{\xi}_2$. This results in expressing (\ref{torusshadow7}) in terms of eight independent terms

\begin{equation} \label{torusshadow12}
\begin{aligned}
W_{\Delta_1, \Delta_2}^{h_{1}, h_{2}}(q, \bar{q}, Z_1, \bar{Z}_1, Z_2, \bar{Z}_2)=q^{\Delta_{1}} \bar{q}^{\Delta_1} & \int d^{2} w_{1} d^{2} w_{2}  \left(f_{1}+  \theta_{1} \theta_{2} f_{2} +\theta_{1} \bar{\theta}_{1} f_{3}+\theta_{2} \bar{\theta}_{2} f_{4}+   \right. \\ &+  \theta_{1} \bar{\theta}_{2} f_{5}+
  \bar{\theta}_{1} \theta_{2} f_{6}+\bar{\theta}_{1} \bar{\theta}_{2} f_{7}+
\left.\theta_{1} \bar{\theta}_{1} \theta_{2} \bar{\theta}_{2} f_{8}\right).
\end{aligned}
\end{equation}
Each term $f_{i}$ can be used to compute superconformal blocks corresponding to different parts of (\ref{thetorus3-1}). Here, we will analyze in detail the terms $f_{1}$ and $f_{2}$. For other terms, we will provide the final result since the analysis follows the same rationale. For the purely bosonic term, i.e., the term $f_1$, we obtain
\begin{equation}\label{torusshadow13}
\begin{aligned}
& f_{1}  =\tilde { C } ^{*}_ {  \Delta_1  h _ { 1 }  \Delta_2 } \tilde { C }^{*} _ {  \Delta_2  h _ { 2 }  \Delta_1 } \left|v_{1-\Delta_1, h_1, \Delta_2} \left(w_{1}, z_{1}, w_{2}\right) v_{1-\Delta_2, h_2, \Delta_1}\left(w_{2}, z_{2}, q w_{1}\right)-  \right. \\
& \left.-\sqrt{q} v_{1 / 2-\Delta_1, h_1, \Delta_2+1 / 2}\left(w_{1}, z_{1}, w_{2}\right) v_{1 / 2-\Delta_2, h_2, \Delta_1+1 / 2}\left(w_{2}, z_{2}, q w_{1}\right)\right|^2 \\
& +C^{*}_{\Delta_1 h_1 \Delta_2} C^{*}_{\Delta_2 h_2\Delta_1} \\
& \times\left|\left(1 / 2-\Delta_1+\Delta_2-h_1\right) v_{1-\Delta_1, h_1, \Delta_2+1 / 2}\left(w_{1}, z_{1}, w_{2}\right) v_{1 / 2-\Delta_2, h_2, \Delta_1}\left(w_{2}, z_{2}, q w_{1}\right)+\right. \\
& \left.-\left(1 / 2+\Delta_1-\Delta_2-h_2\right) \sqrt{q} v_{1 / 2-\Delta_1, h_1, \Delta_2 }\left(w_{1}, z_{1}, w_{2}\right) v_{1-\Delta_2, h_2, \Delta_1+1 / 2}\left(w_{2}, z_{2}, q w_{1}\right)\right|^2.
\end{aligned}
\end{equation}
Since the structure constants  $\tilde { C } ^{*}_ {  \Delta_i  h _ { j }  \Delta_k }, C ^* _ {  \Delta_i  h _ { j }  \Delta_k } $ are independent, $f_{1}$ consists of two independent terms, each providing different conformal blocks. Focusing on the holomorphic part of $f_{1}$ and taking the integral as discussed for the $\mathfrak{sl}(2)$ case, we obtain the integral representation for $B^{(1)}_{00}$ from the term proportional to $\tilde { C } ^{*}_ {  \Delta_1  h _ { 1 }  \Delta_2 } \tilde { C }^{*} _ {  \Delta_2  h _ { 2 }  \Delta_1 } $, as follows
\begin{equation}  \label{torusshadow13-1}
\begin{aligned}
 &   B^{(1)}_{00} \left(q,z_1, z_2\right)  \\&  =\frac{q^{\Delta_1}}{c_2\left(\text{\small$h_1, h_2,\Delta_1, \Delta_2$}\right)}\int_{\mathbf{C}_1} dw_1\int_{\mathbf{C}_2} dw_2  \bigg(  v_{1-\Delta_1, h_1, \Delta_2} (w_{1}, z_{1}, w_{2}) v_{1-\Delta_2, h_2, \Delta_1}(w_{2}, z_{2}, q w_{1})    \\
& -\sqrt{q} v_{1 / 2-\Delta_1, h_1, \Delta_2+1 / 2}\left(w_{1}, z_{1}, w_{2}\right) v_{1 / 2-\Delta_2, h_2, \Delta_1+1 / 2}\left(w_{2}, z_{2}, q w_{1}\right) \bigg) \\&
    = \mathcal{F}_{\Delta_1, \Delta_2}^{h_1, h_2}(q, z_1, z_2) \\ &-\frac{(\Delta_1+ \Delta_2-h_1) (\Delta_1+\Delta_2-h_2) }{4 \Delta_1 \Delta_2}  \mathcal{F}_{\Delta_1+1/2, \Delta_2+1/2}^{h_1, h_2}(q, z_1, z_2).
\end{aligned}
\end{equation} 
Similarly, from the term proportional to $C^*_{\Delta_1 h_1 \Delta_2} C^*_{\Delta_2 h_2\Delta_1}$, we obtain the integral representation for $B^{(2)}_{00}$:
\begin{equation} \label{torusshadow14}
\begin{aligned}
&  B^{(2)}_{00} \left(q,z_1, z_2\right)  = \frac{q^{\Delta_1}}{c_2\left(\text{\small$h_1, h_2,\Delta_1, \Delta_2+1/2$}\right)}  \\&  \times  \int_{\mathbf{C}_1} dw_1\int_{\mathbf{C}_2} dw_2 \bigg(  v_{1-\Delta_1, h_1, \Delta_2+1 / 2}\left(w_{1}, z_{1}, w_{2}\right) v_{1 / 2-\Delta_2, h_2, \Delta_1}\left(w_{2}, z_{2}, q w_{1}\right)   \\&  - \frac{\left(1 / 2+\Delta_1-\Delta_2-h_2\right)}{ \left(1 / 2-\Delta_1+\Delta_2-h_1\right) } \sqrt{q} v_{1 / 2-\Delta_1, h_1, \Delta_2 }\left(w_{1}, z_{1}, w_{2}\right) v_{1-\Delta_2, h_2, \Delta_1+1 / 2}\left(w_{2}, z_{2}, q w_{1}\right)  \bigg)  \\
& =\mathcal{F}_{\Delta_1, \Delta_2+1 / 2}^{h_1, h_2}\left(q, z_{1}, z_{2}\right)  -\frac{\Delta_2}{\Delta_1} \mathcal{F}_{\Delta_1+1 / 2, \Delta_2}^{h_1, h_2}\left(q, z_{1}, z_{2}\right) .
\end{aligned}
\end{equation}
Notice that in the limit $h_2 \rightarrow 0, \Delta_2 \rightarrow \Delta_1$, the expression (\ref{torusshadow13-1}) reduces to $B_0$ from (\ref{thetorus5-2}), which is the desired relation for the purely bosonic two-point torus superconformal block. For (\ref{torusshadow14}), such a limit cannot be imposed since $\Delta_1, \Delta_2$ differ by $1 / 2$ in both terms on r.h.s of (\ref{torusshadow14}).  

\hfill\break
\textbf{Contribution $\theta_{1} \theta_{2}$} : Now we proceed with the term $f_2$ from (\ref{torusshadow12}). This term is given by 

\begin{equation}\label{torusshadow15}
\begin{aligned}
& f_{2}\big|_{\bar{q}\to 0} \\&=\tilde{C}^*_{\Delta_1 h_1 \Delta_2} \tilde{C}^*_{\Delta_2 h_2 \Delta_1}\left(\tilde{a}_{6}^{(1)} v_{1-\Delta_1, h_1+1 / 2, \Delta_2+1 / 2}\left(w_{1}, z_{1}, w_{2}\right) v_{1 / 2-\Delta_2, h_2+1 / 2, \Delta_1}\left(w_{2}, z_{2}, q w_{1}\right)\right. \\
& \left.-\tilde{a}_{6}^{(2)}\sqrt{q} v_{1 / 2-\Delta_1, h_1+1 / 2, \Delta_2}\left(w_{1}, z_{1}, w_{2}\right) v_{1-\Delta_2, h_2+1 / 2, \Delta_1+1 / 2}\left(w_{2}, z_{2}, q w_{1}\right)\right) \\
& \times\left(\bar{v}_{1-\Delta_{1}, h_{1}, \Delta_{2}}\left(\bar{w}_{1}, \bar{z}_{1}, \bar{w}_{2}\right) \bar{v}_{1-\Delta_{2}, h_{2}, \Delta_{1}}\left(\bar{w}_{2}, \bar{z}_{2}, 0\right)\right) \\
& +C^*_{\Delta_1 h_1 \Delta_2} C^*_{\Delta_2 h_2 \Delta_1}\left(a_{6}^{(1)} v_{1-\Delta_1, h_1+1 / 2, \Delta_2}\left(w_{1}, z_{1}, w_{2}\right) v_{1-\Delta_2, h_2+1 / 2, \Delta_1}\left(w_{2}, z_{2}, q w_{1}\right)\right. \\
& \left.+a_{6}^{(2)} \sqrt{q} v_{1 / 2-\Delta_1, h_1+1 / 2, \Delta_2+1 / 2}\left(w_{1}, z_{1}, w_{2}\right) v_{1 / 2-\Delta_2, h_2+1 / 2, \Delta_1+1 / 2}\left(w_{2}, z_{2}, q w_{1}\right)\right) \\
& \times\left(\bar{v}_{1-\Delta_{1}, h_{1}, \Delta_{2}+1 / 2}\left(\bar{w}_{1}, \bar{z}_{1}, \bar{w}_{2}\right) \bar{v}_{1 / 2-\Delta_{2},h_{2}, \Delta_{1}}\left(\bar{w}_{2}, \bar{z}_{2}, 0\right)\right) ,
\end{aligned}
\end{equation}
where we used
\begin{equation} \label{torusshadow16}
    \begin{array}{ll}
   \tilde{a}_{6}^{(1)}=2\left(-\Delta_1+\Delta_2+h_1\right) ,     &   a_{6}^{(1)}=-(-\Delta_1-\Delta_2+h_1+\frac{1}{2} )(-\Delta_1-\Delta_2+h_2+\frac{1}{2}),\\
      \tilde{a}_{6}^{(2)}=2\left(\Delta_1-\Delta_2+h_2\right)  ,   &  a_{6}^{(2)}=\left(\Delta_1+\Delta_2+h_1-\frac{1}{2}\right)\left(\Delta_1+\Delta_2+h_2-\frac{1}{2}\right).
    \end{array}
\end{equation}
From the holomorphic part of (\ref{torusshadow15}) we obtain the integral expressions for (\ref{thetorus12}, \ref{thetorus13}). Thus, from the term proportional to $\tilde{C}^*_{\Delta_1 h_1 \Delta_2} \tilde{C}^*_{\Delta_2 h_2 \Delta_1}$, we obtain

\begin{equation} \label{torusshadow17}
\begin{aligned}
&  B_{\theta_{1} \theta_{2}}^{(1)}=  \frac{q^{\Delta_1}}{ c_2\left(\text{\small$h_1+1/2, h_2+1/2,\Delta_1, \Delta_2+1/2$}\right)} \\& \times\int_{\mathbf{C}_{1}} d w_{1} \int_{\mathbf{C}_{2}} d w_{2}\bigg( v_{1-\Delta_1, h_1+1 / 2, \Delta_2+1 / 2}\left(w_{1}, z_{1}, w_{2}\right) v_{1 / 2-\Delta_2, h_2+1 / 2, \Delta_1}\left(w_{2}, z_{2}, q w_{1}\right)\\&  - \frac{\tilde{a}_{6}^{(2)} }{\tilde{a}_{6}^{(1)}}\sqrt{q}  v_{1 / 2-\Delta_1, h_1+1 / 2, \Delta_2}\left(w_{1}, z_{1}, w_{2}\right) v_{1-\Delta_2, h_2+1 / 2, \Delta_1+1 / 2}\left(w_{2}, z_{2}, q w_{1}\right)\bigg)\\&=
\mathcal{F}_{\Delta_1, \Delta_2+1 / 2}^{h_1+1 / 2, h_2+1 / 2}\left(q, z_{1}, z_{2}\right)-\frac{\Delta_2\left(\Delta_1-\Delta_2+h_{1}\right)\left(\Delta_2-\Delta_1-h_{2}\right)}{\Delta_1\left(\Delta_1-\Delta_2-h_{1}\right)\left(\Delta_2-\Delta_1+h_{2}\right)} \mathcal{F}_{\Delta_1+1 / 2, \Delta_2}^{h_1+1 / 2, h_2+1 / 2}\left(q, z_{1}, z_{2}\right),
\end{aligned}
\end{equation}
and similarly, from the term proportional to $C^*_{\Delta_1 h_1 \Delta_2} C^*_{\Delta_2 h_2 \Delta_1}$, we obtain the integral representation for $  B_{\theta_{2} \theta_{1}}^{(2)}$
\begin{equation}
    \begin{aligned}
    &    B_{ \theta_{1} \theta_{2}}^{(2)}= \frac{q^{\Delta_1}}{c_2\left(\text{\small$h_1+1/2, h_2+1/2,\Delta_1, \Delta_2$}\right)}   \\& \times \int_{\mathbf{C}_{1}} d w_{1} \int_{\mathbf{C}_{2}} d w_{2}\bigg( v_{1-\Delta_1, h_1+1 / 2, \Delta_2}\left(w_{1}, z_{1}, w_{2}\right) v_{1-\Delta_2, h_2+1 / 2, \Delta_1}\left(w_{2}, z_{2}, q w_{1}\right) \\&  +\frac{a_{6}^{(2)} }{a_{6}^{(1)} } \sqrt{q} v_{1 / 2-\Delta_1, h_1+1 / 2, \Delta_2+1 / 2}\left(w_{1}, z_{1}, w_{2}\right) v_{1 / 2-\Delta_2, h_2+1 / 2, \Delta_1+1 / 2}\left(w_{2}, z_{2}, q w_{1}\right)\bigg) \\&=
    \mathcal{F}_{\Delta_1, \Delta_2}^{h_1+1 / 2, h_2+1 / 2}\left(q, z_{1}, z_{2}\right)-\alpha_{3} \mathcal{F}_{\Delta_1+1 / 2, \Delta_2+1 / 2}^{h_1+1 / 2, h_2+1 / 2}\left(q, z_{1}, z_{2}\right),
    \end{aligned}
\end{equation}
where

\begin{equation} \label{torusshadow19}
\alpha_{3}=\frac{\left(2 \Delta_1+2 \Delta_2+2 h_1-1\right)\left(2 \Delta_1+2 \Delta_2+2 h_2-1\right)}{16 \Delta_1 \Delta_2}.
\end{equation}
In the next section, it will be necessary to work with the redefined $B_{\theta_1 \theta_2}^{(1), (2)}$  obtained by multiplying them by the following constants

\begin{equation} \label{torusshadow19-1}
\begin{aligned}
& \tilde{B}_{\theta_{1} \theta_{2}}^{(1)}=\alpha_{4} B_{\theta_{1} \theta_{2}}^{(1)}, \\
& \widetilde{B}_{\theta_{1} \theta_{2}}^{(2)}=\alpha_{5} B_{\theta_{1} \theta_{2}}^{(2)},
\end{aligned}
\end{equation}
where

\begin{equation} \label{torusshadow19-2}
\begin{aligned}
& \alpha_{4}=\frac{\left(\Delta_{1}-\Delta_{2}-h_{1}\right)\left(-\Delta_{1}+\Delta_{2}+h_{2}\right)}{2 \Delta_{2}} ,\\
& \alpha_{5}=-2 \Delta_{2}.
\end{aligned}
\end{equation}
The constants (\ref{torusshadow19-2}) arise when considering equation (\ref{thetorus18-0}) and choosing the same normalization for equations (\ref{thetorus12}) and (\ref{thetorus13}) (the denominators of those equations) as for equations (\ref{thetorus9}) and (\ref{thetorus10}), respectively.

We can repeat the same procedure for the other terms $f_{i}$ of (\ref{torusshadow12}). The results we obtained for the other holomorphic superconformal blocks are listed as follows:

\hfill\break
\textbf{Contribution $\theta_{1} \bar{\theta}_{1}$ :} From the term $f_3$ we obtain the superconformal blocks

\begin{equation} \label{torusshadow20}
\begin{aligned}
& B_{\theta_{1} \bar{\theta}_{1}}^{(1)}=\mathcal{F}_{\Delta_1, \Delta_2}^{h_1+1 / 2, h_2}\left(q, z_{1}, z_{2}\right) \\
& -\frac{\left(-1+2 \Delta_1+2 \Delta_2+2 h_1\right)\left(\Delta_1+\Delta_2-h_2\right)}{8 \Delta_1 \Delta_2} \mathcal{F}_{\Delta_1+1 / 2, \Delta_2+1 / 2}^{h_1+1 / 2, h_2}\left(q, z_{1}, z_{2}\right),
\end{aligned}
\end{equation}
\begin{equation}
    \begin{aligned}
        B_{\theta_{1} \bar{\theta}_{1}}^{(2)}=& \mathcal{F}_{\Delta_1, \Delta_2+1 / 2}^{h_1+1 / 2, h_2}\left(q, z_{1}, z_{2}\right) \\&
-\frac{\Delta_2\left(\Delta_1-\Delta_2+h_1\right)}{\Delta_1\left(\Delta_1-\Delta_2-h_1\right)} \mathcal{F}_{\Delta_1+1 / 2, \Delta_2}^{h_1+1 / 2,h_2}\left(q, z_{1}, z_{2}\right).
    \end{aligned}
\end{equation}
In the limit $h_2 \rightarrow 0, \Delta_2 \rightarrow \Delta_1$, the expression (\ref{torusshadow20}) reduces to $B_1$ from (\ref{thetorus5-2}), which is the desired relation.

\hfill\break
\textbf{Contribution $ \theta_2 \bar{\theta}_2 $:} From the term $f_4$ we obtain the superconformal blocks

\begin{equation}
\begin{aligned}
   B^{(1)}_{  \theta_2 \bar{\theta}_2 }=     & \mathcal{F}_{\Delta_1,\Delta_2}^{h_1,h_2+\frac{1}{2}}  \left(q, z_{1}, z_{2}\right)\\&-\frac{\left(\Delta_1+\Delta_2-h_1\right) \left(2 \Delta_1+2 \Delta_2+2 h_2-1\right)}{8 \Delta_1 \Delta_2}\mathcal{F}_{\Delta_1+\frac{1}{2},\Delta_2+\frac{1}{2}}^{h_1,h_2+\frac{1}{2}} \left(q, z_{1}, z_{2}\right),
\end{aligned}
\end{equation}
\begin{equation}
   B^{(2)}_{  \theta_2 \bar{\theta}_2 }= \mathcal{F}_{\Delta_1,\Delta_2+\frac{1}{2}}^{h_1,h_2+\frac{1}{2}} \left(q, z_{1}, z_{2}\right) -\frac{\Delta_2\left(\Delta_1-\Delta_2+h_2\right) }{\Delta_1 \left(\Delta_1-\Delta_2-h_2\right)}\mathcal{F}_{\Delta_1+\frac{1}{2},\Delta_2}^{h_1,h_2+\frac{1}{2}} \left(q, z_{1}, z_{2}\right).
\end{equation}

\hfill\break
\textbf{Contribution $\theta_{1} \bar{\theta}_{2}$}: From the term $f_5$ we obtain the superconformal blocks

\begin{equation}
    \begin{aligned}
        B_{\theta_{1} \bar{\theta}_{2}}^{(1)}=     B_{\theta_{1} \bar{\theta}_{1}}^{(1)},
    \end{aligned}
\end{equation}
\begin{equation} \label{torusshadow21}
\begin{gathered}
B_{\theta_{1} \bar{\theta}_{2}}^{(2)}=     B_{\theta_{1} \bar{\theta}_{1}}^{(2)} .
\end{gathered}
\end{equation}

\hfill\break
\textbf{Contribution $\bar{\theta}_{1} \theta_{2}$:} From the term $f_6$ we obtain the superconformal blocks

\begin{equation}
    \begin{aligned}
        B_{\bar{\theta}_{1} \theta_{2}}^{(1)}=      B_{\theta_{2} \bar{\theta}_{2}}^{(1)},
    \end{aligned}
\end{equation}
\begin{equation} \label{torusshadow22}
\begin{gathered}
B_{\bar{\theta}_{1} \theta_{2}}^{(2)}=     B_{\theta_{2} \bar{\theta}_{2}}^{(2)}.
\end{gathered}
\end{equation}

\hfill\break
\textbf{Contribution $ \bar{\theta}_1 \bar{\theta}_2 $:} From the term $f_7$ we obtain the superconformal blocks
\begin{equation}
\begin{aligned}
      &  B^{(1)}_{\bar{\theta}_1 \bar{\theta}_2} =B^{(1)}_{00},
       \end{aligned}
\end{equation}
\begin{equation}
    \begin{aligned}
          B^{(2)}_{\bar{\theta}_1 \bar{\theta}_2} =B^{(2)}_{00}.
    \end{aligned}
\end{equation}

\hfill\break
\textbf{Contribution $ \theta_1 \bar{\theta}_1  \theta_2 \bar{\theta}_2 $:} From the term $f_8$ we obtain the superconformal blocks
\begin{equation}
\begin{aligned}
      &  B^{(1)}_{ \theta_1 \bar{\theta}_1  \theta_2 \bar{\theta}_2} =B^{(1)}_{\theta_1 \theta_2},
       \end{aligned}
\end{equation}
\begin{equation}
    \begin{aligned}
               B^{(2)}_{ \theta_1 \bar{\theta}_1  \theta_2 \bar{\theta}_2} =B^{(2)}_{\theta_1 \theta_2}.
    \end{aligned}
\end{equation}
\section{Casimir Operator for superconformal blocks} \label{sec:casimirini}
In this section, we will check that the superconformal blocks (\ref{torusshadow13-1}, \ref{torusshadow14}  \ref{torusshadow19-1}) satisfy differential equations derived from the Casimir operator.
First, let us derive the differential equations. The $\mathfrak{osp}(1|2)$ Casimir operator is given by

\begin{equation} \label{casimir1}
S_{2}=-L_{0}^{2}+\frac{1}{2}\left(L_{1} L_{-1}+L_{-1} L_{1}\right)+\frac{1}{4}\left(G_{-1 / 2} G_{1 / 2}-G_{1 / 2} G_{-1 / 2}\right).
\end{equation}
One can insert the Casimir operator in the following two ways

\begin{equation} \label{casimir2}
\begin{array}{r}
\operatorname{str}_{\Delta_{1}}\left[S_{2} q^{L_{0}} \phi_{h_{1}} \mathbb{P}_{\Delta_2} \phi_{h_{2}}\right]=-\Delta_{1}\left(\Delta_{1}-\frac{1}{2} \right) \operatorname{str}_{\Delta_{1}}\left[q^{L_{0}} \phi_{h_{1}} \mathbb{P}_{\Delta_2}\phi_{h_{2}}\right] ,
\end{array}
\end{equation}
\begin{equation}  \label{casimir2-1}
    \operatorname{str}_{\Delta_{1}}\left[q^{L_{0}} \phi_{h_{1}} S_{2} \mathbb{P}_{\Delta_2}\phi_{h_{2}}\right]=-\Delta_{2}\left(\Delta_{2}-\frac{1}{2}\right )  \operatorname{str}_{\Delta_{1}}\left[q^{L_{0}} \phi_{h_{1}} \mathbb{P}_{\Delta_2}\phi_{h_{2}}\right].
\end{equation}
For each particular insertion, we obtain different eigenvalues as described by the r.h.s of the above equations. Here, $\mathbb{P}_{{\Delta}_2}$ stands for the projector onto $V_{\Delta_2}$. Next, we write the l.h.s of (\ref{casimir2}, \ref{casimir2-1}) as differential operators. We do this using the standard procedure described in \cite{Kraus:2017ezw, Alkalaev2018, Alkalaev2022}. For this, we require the following commutation and anticommutation relations

\begin{equation} \label{casimir3}
\begin{aligned}
& {\left[L_{m}, \phi_{h_i}(z_i)\right]=\mathcal{L}_{m}^{(i)} \phi_{h_i}(z_i)}, \quad\mathcal{ L}^{(i)}_m= z_i^{m}\left(z_i \partial_{z_i}+(m+1) h_i \right),\\
& {\left[G_{r}, \phi_{h_i}(z_i)\right]=z_i^{r+1/2 }\psi_{h_i}(z_i)}, \\
& {\left[L_{m}, \psi_{h_i}(z_i)\right]=z_i^{m}(z_i \partial_{z_i}+(m+1)(h_i+\frac{1}{2})) \psi_{h_i}(z_i)} ,\\
& \left\{G_{r}, \psi_{h_i}(z_i)\right\}=\mathcal{G}_r^{(i)} \phi_{h_i}(z_i),\quad \mathcal{G}_r^{(i)}= z_i^{r-\frac{1}{2}}\left(z_i \partial_{z_i}+(2 r+1) h_i\right).
\end{aligned}
\end{equation}
The simplest term of the Casimir operator to treat is $L_{0}^{2}$, for which we have

\begin{equation} \label{casimir4}
 \operatorname{str}_{\Delta_{1}}\left[L_{0}^{2} q^{L_{0}} \phi_{h_1} \phi_{h_{2}}\right]=\left(q \partial_{q}\right)^{2} \operatorname{str}_{\Delta_{1}}\left[q^{L_{0}} \phi_{h_{1}} \phi_{h_{2}}\right].
\end{equation}
The insertions of other operators, e.g., $L_{0}^{2}, L_{1} L_{-1}$, can be computed by moving these operators to the most-right side of the trace (this is done by using the relations (\ref{casimir3})) and then using the graded cyclic property of the supertrace $\operatorname{str}_{ \Delta_1}$. One can show that these insertions result in the equations

\begin{equation} \label{casimir5}
\operatorname{str}_{\Delta_{1}}\left[q^{L_{0}} \phi_{h_{1}} L_{0}^{2} \phi_{h_{2}}\right]=\left(\left(\mathcal{L}_{0}^{(2)}\right)^{2}+2 \mathcal{L}_{0}^{(2)} q \partial_{q}+\left(q \partial_{q}\right)^{2}\right)  \operatorname{str}_{\Delta_{1}}\left[q^{L_{0}} \phi_{h_{1}} \phi_{h_{2}}\right],
\end{equation}
and

\begin{equation} \label{casimir6}
 \operatorname{str}_{\Delta_{1}}\left[L_{1} L_{-1} q^{L_{0}} \phi_{h_{1}} \phi_{h_{2}}\right]=\left(\frac{2}{1-q} q \partial_{q}-\frac{q}{(1-q)^{2}} \hat{A}_{1}\right) \operatorname{str}_{\Delta_{1}}\left[q^{L_0} \phi_{h_{1}} \phi_{h_{2}}\right],
\end{equation}

\begin{equation} \label{casimir7}
\operatorname{str}_{\Delta_{1}}\left[q^{L_{0}} \phi_{h_{1}} L_{1} L_{-1} \phi_{h_{2}}\right]=\left(\frac{2}{1-q} q \partial_{q}+\frac{2}{1-q} \mathcal{L}_{0}^{(2)}-\frac{1}{(1-q)^{2}} \hat{A}_{2}\right)  \operatorname{str}_{\Delta_{1}}\left[q^{L_{0}} \phi_{h_{1}} \phi_{h_{2}}\right],
\end{equation}
where we used
\begin{equation} \label{casimir8}
\begin{aligned}
& \hat{A}_1=\left(\mathcal{L}_{-1}^{(1)}+\mathcal{L}_{-1}^{(2)}\right)\left(\mathcal{L}_{1}^{(1)}+\mathcal{L}_{1}^{(2)}\right) ,\\
& \hat{A}_{2}=\left(\mathcal{L}_{-1}^{(1)}+q \mathcal{L}_{-1}^{(2)}\right)\left(\mathcal{L}_{1}^{(2)}+q \mathcal{L}_{1}^{(1)}\right).
\end{aligned}
\end{equation}
The insertion of the operators $G_{-1 / 2} G_{1 / 2}$ in (\ref{casimir2}, \ref{casimir2-1}) generates also terms proportional to $\operatorname{str}_{\Delta_{1}}\left[q^{L_{0}} \psi_{h_{1}} \psi_{h_{2}}\right]$. By using the same idea used for generators $L_{i}$, one obtains the following relations

\begin{equation} \label{casimir9}
\begin{aligned}
& \operatorname{str}_{\Delta_{1}}\left[G_{-r} G_{r} q^{L_{0}} \phi_{h_{1}} \phi_{h_{2}}\right]= \\& \left(\frac{2 q \partial_{q}}{\left(1-q^{-r}\right)}+\frac{q^{r}}{\left(1-q^{r}\right)^{2}}\left(z_{1}^{-r+\frac{1}{2}} \mathcal{G}_{r}^{(1)}+z_{2}^{-r+\frac{1}{2}} \mathcal{G}_{r}^{(2)}\right) \right) \operatorname{str}_{\Delta_{1}}\left[q^{L_{0}} \phi_{h_{1}} \phi_{h_{2}}\right] \\
& +\frac{q^{r}}{\left(1-q^{r}\right)^{2}}\left(z_{1}^{r+\frac{1}{2}} z_{2}^{-r+\frac{1}{2}}-z_{1}^{-r+\frac{1}{2}} z_{2}^{r+\frac{1}{2}}\right) \operatorname{str}_{\Delta_{1}}\left[q^{L_{0}} \psi_{h_{1}} \psi_{h_{2}}\right],
\end{aligned}
\end{equation}
and

\begin{equation} \label{casimir10}
\begin{aligned}
& \operatorname{str}_{\Delta_{1}}\left[q^{L_{0}} \phi_{h_{1}} G_{\frac{1}{2}} G_{-\frac{1}{2}} \phi_{h_{2}}\right]=\\& \left(\frac{2( \mathcal{L}_0^{(2)}+q \partial q)}{\left(1-q^{\frac{1}{2}}\right)}+\frac{q^{\frac{1}{2}}}{\left(1-q^{\frac{1}{2}}\right)^{2}}\left(z_{2} \partial_{z_{2}}+z_{1} \partial_{z_{1}}\right)\right) \operatorname{str}_{\Delta_{1}}\left[q^{L_{0}} \phi_{h_{1}} \phi_{h_{2}}\right] \\
& +\frac{1}{\left(1-q^{\frac{1}{2}}\right)^{2}}\left(z_{2}-q z_{1}\right)  \operatorname{str}_{\Delta_{1}}\left[q^{L_{0}} \psi_{h_{1}} \psi_{h_{2}}\right].
\end{aligned}
\end{equation}
From Ward's identity, we also have

\begin{equation} \label{casimir11}
\left(z_{1} \partial_{z_{1}}+z_{2} \partial z_{2}\right)  \operatorname{str}_{\Delta_{1}}\left[q^{L_{0}} \phi_{h_{1}} \phi_{h_{2}}\right]=-\left(h_{1}+h_{2}\right)  \operatorname{str}_{\Delta_{1}}\left[q^{L_{0}} \phi_{h_{1}} \phi_{h_{2}}\right].
\end{equation}
By substituting (\ref{casimir4}, \ref{casimir5}, \ref{casimir6}, \ref{casimir7}, \ref{casimir9}, \ref{casimir10}) into (\ref{casimir2}, \ref{casimir2-1}), using (\ref{casimir11}) and writing the resulting equations in components, we obtain from (\ref{casimir2}) the first differential equation for the superconformal blocks

\begin{equation}\label{casimir12}
\begin{aligned}
& \left(-q \partial_{q}-q^{2} \partial_{q}^{2}+\frac{q(1-q^{1 / 2})}{2(1+q^{1/2})} \partial_{q}-\frac{q}{(1-q)^{2}} \hat{A}_1+\frac{q^{1 / 2}}{2(1-q^{1/2})^{2}}(h_{1}+h_{2})     \right. \\ & \quad + \left.   \Delta_{1}  (\Delta_{1}-\frac{1}{2})\right) B_{00}^{(i)}+\frac{q^{1 / 2}}{2(1-q^{1/2})^{2}}\left(z_{1}-z_{2}\right) \tilde{B}_{\theta_{1} \theta_{2}}^{(i)}=0  ,
\end{aligned}
\end{equation}
and from (\ref{casimir2-1}), we obtain the second differential equation

\begin{equation} \label{casimir13}
    \begin{aligned}
        & \left(-q^{2} \partial_{q}^{2}+\frac{2 q^{2}}{1-q} \partial_{q}-\frac{1}{(1-q)^2}\hat{A}_{2}+\frac{(1+q)}{ 1-q}  \mathcal{L}_{0}^{(2)}- \left((\mathcal{L}_{0}^{(2)})^{2}+2 q \mathcal{L}_{0}^{(2)} \partial_{q} \right) \right. \\
& +\Delta_{2}(\Delta_{2}-1 / 2)+\frac{1}{2}(\mathcal{L}_{0}^{(2)}+q \partial_{q})-\frac{1}{1-q^{1/2}}(\mathcal{L}_{0}^{(2)}+q \partial_{q}) \\
& \left.+\frac{q^{\frac{1}{2}} }{2 (1-q^{1/2})^2}\left(h_{1}+h_{2}\right)\right) B_{00}^{(i)}  -\frac{1}{2 (1-q^{1/2})^2}\left(z_{2}-q z_{1}\right) \tilde{B}_{\theta_{1} \theta_{2}}^{(i)} =0  .
    \end{aligned}
\end{equation}
where $i=1,2$. It is straightforward to verify these differential equations perturbatively by expanding the superconformal blocks in $z_2/z_1$ and $q$. For this type of expansion, it is convenient to use the representation (\ref{ap:2ptgcb10}) for $\mathcal{F}$. One can also prove equations (\ref{casimir12}, \ref{casimir13}) exactly. For the general proof, it is convenient to use the representation (\ref{torusshadow4}). The proof can be summarized in three key steps, outlined below:

\hfill\break
\textbf{Step 1}. We rewrite (\ref{casimir12}, \ref{casimir13}) in terms of $\partial_{q}, \partial_{\rho_1}, \partial_{\rho_2}$. Using the representation (\ref{torusshadow4}), we substitute (\ref{torusshadow13-1}, \ref{torusshadow14}  \ref{torusshadow19-1}) into (\ref{casimir12}, \ref{casimir13}). After this substitution, we get rid of the overall factor in front of Appell function $F_4$ in (\ref{torusshadow4}), obtaining that (\ref{casimir12}, \ref{casimir13}) become differential equations for functions $F_4$, each of the obtained differential equation involves four different functions $F_4$ (this is so, because, $B_{00}^{(i)}$ and $\tilde{B}_{\theta_1 \theta_2}^{(i)}$ are given by combinations of two Appell functions $F_4$ with different arguments). This step can be performed straightforwardly.

\hfill\break
\textbf{Step 2}. We rewrite the differential equations obtained in the previous step entirely in terms of the variables $\rho_{1}, \rho_{2}$. This is achieved by splitting each differential equation into two terms: one containing only integer powers of $q$ and the other containing only half-integer powers of $q$. Each of these terms can then be written as a differential equation involving only $\rho_{1}, \rho_{2}$. Due to the non simple relations (\ref{ap:2ptgcb6}) between $z_1, z_2, q$, and $\rho_1, \rho_2$, this task  turns out to be intricate. Even though the expressions obtained do not simplify in a simple way, the change of variables can be performed.

\hfill\break
\textbf{Step 3}. In the final step, we verify that the differential equations obtained in the previous step are all satisfied. We do this by converting the differential equations into recurrence relations for the series coefficients of the functions $F_4$.

\section{Conclusion and outlook} \label{sec:conclusions}
In this work, we have generalized the shadow formalism to $N=1$ two-dimensional superconformal field theory in the Neveu-Schwarz sector and used it to compute global $\mathfrak{osp}(1|2)$  superconformal blocks, which arise in the large central charge limit of the superconformal theory. An essential ingredient of the shadow formalism is the so-called shadow operator that we have explicitly constructed in (\ref{superShadowDef}) for the scalar superfields. We demonstrated that the two-point function of a superfield with its shadow factorizes into a product of delta functions (\ref{SuperDelta}) of spatial and Grassmann coordinates, and therefore the projector-like operator (\ref{projectorDef}) can be constructed. The shadow operator allows us to construct an identity-like operator (\ref{SuperIdentity}), which can be used to decompose correlation functions of superfields into superconformal partial waves.
In sections \ref{CBExamples} and \ref{sec:torusbsh1}, we have applied this formalism to compute the four-point superconformal block on a plane, and one- and two-point blocks on a torus.
For four-point spherical superconformal blocks and one-point torus superconformal blocks, our results agree with earlier results obtained by other methods. For two-point torus superconformal blocks, we have verified that the expressions obtained via shadow formalism satisfy required nontrivial relations for torus superconformal blocks. In particular, in section \ref{sec:casimirini}, we showed that the two-point torus superconformal blocks (involving bosonic $\phi_{h_i}$ and fermionic $\psi_{h_i}$ components) satisfy the differential equations which follow from the $\mathfrak{osp}(1|2)$ Casimir operator. These results show that the constructed supersymmetric shadow formalism provides correct integral representations of superconformal blocks both on a plane and on a torus.

There are several related problems that we plan to explore. 
To have the complete picture about $N=1$ supersymmetry theory we need to consider the shadow formalism in the Ramond sector. From the holography perspective, it would be interesting to see the explicit interpretation of the global higher-point superconformal blocks as dual geodesic diagrams.
There are questions about shadow formalism relevant beyond the supersymmetric case. For instance, 
whether the shadow formalism can be generalized to the full Virasoro algebra, enabling us to go beyond the semiclassical limit. 

\acknowledgments
We thank Mikhail Pavlov for fruitful discussions.

\appendix
\section{Calculation of the four-point superconformal block} \label{ap: calculation4ptsbs}
In the section \ref{CBExamples}, we have computed the component $g^{(0,0)}_h$ of the four-point superconformal block.
Below we demonstrate how the rest can be computed using the superanalyticity of the correlation function. The component $g^{(1,0)}_h(X,\bar{X})$ of the superconformal block (\ref{4ptExpansion}) can be computed by setting
\be
\label{g10pointsfix}
    \theta_1=\theta_2=\theta_3=0\,,\quad
    \bar{\theta}_1=\bar{\theta}_2=\bar{\theta}_3=0\,,
\ee
which implies
\be
    X=x\,,\quad \eta=z_{12}^{\frac{1}{2}}z_{14}^{-\frac{1}{2}}z_{24}^{-\frac{1}{2}}\theta_4\,,\quad \eta^{\prime}=0\,.
\ee
The conformal partial wave, in this case, evaluates to
\be
 \begin{split}  \Psi^{h_1,\dots,h_4}_h&(z_1,\dots,Z_4;\bar{z}_1,\dots,\bar{Z}_4)=
\Psi^{h_1,\dots,h_4}_h(z_1,\dots,z_4;\bar{z}_1,\dots,\bar{z}_4)\\
    &+\theta_4\bar{\theta_4}(h^{*}-h_{34})^2\int d^2 z_0 
    \frac{C_{h_1h_2h}C_{hh_3h_4}^{*}|z_{12}|^{2h-2h_1-2h_2}|z_{34}|^{1-2h-2h_3-2h_4}}{
    |z_{10}|^{2p_1}|z_{20}|^{2p_2}
    |z_{03}|^{2p_3}|z_{04}|^{2p_4+2}}\\
     &+\theta_4\bar{\theta_4}\int d^2 z_0 
    \frac{\tilde{C}_{h_1h_2h}\tilde{C}_{hh_3h_4}^{*}|z_{12}|^{2h-2h_1-2h_2+1} |z_{34}|^{-2h-2h_3-2h_4}}{|z_{10}|^{1+2p_1}|z_{20}|^{1+2p_2}|z_{03}|^{2p_3-1}|z_{04}|^{1+2p_4}}\,.
\end{split}
\ee
Evaluating the integrals, we get
\be
\begin{split}   \Psi^{h_1,\dots,h_4}_h&(z_1,\dots,Z_4;\bar{z}_1,\dots,\bar{Z}_4)=
\Psi^{h_1,\dots,h_4}_h(z_1,\dots,z_4;\bar{z}_1,\dots,\bar{z}_4)\\
&+\eta\bar{\eta} \mathcal{L}_{h_1,h_2,h_3,h_4}
(h^{*}-h_{34})^2 C_{h_1h_2h}C^{*}_{hh_3h_4}\hgInt\left(h,h_{12},h_{34}-\frac{1}{2}\Big|X,\bar{X}\right)\\
    &+\eta\bar{\eta} \mathcal{L}_{h_1,h_2,h_3,h_4}\tilde{C}_{h_1h_2h}\tilde{C}^{*}_{hh_3h_4}
    \hgInt\left(h+\frac{1}{2},h_{12},h_{34}-\frac{1}{2}\Big|X,\bar{X}\right)\,.
\end{split}
\ee
Only the last two terms contribute to $g^{(1,0)}_h$. Moreover, there are no other contributions, as the factor $\theta_4\bar{\theta}_4$ couldn't have arisen from the expansion of the leg factor, as the holomorphic Grassmann variables in the expansion of the latter always come in pairs, i.e. $\theta_i\theta_j$, but all such pairs vanish at the chosen point (\ref{g10pointsfix}) of the superspace.
Separating the contributions of the global conformal block and that of the shadow global conformal block, we obtain the following expression for the function $g^{(1,0)}_h$:
\be
    g^{(1,0)}_h(X,\bar{X})=\tilde{C}_{hh_1h_2}C_{hh_3h_4}|G^{(o)}_{1,0}(h,h_{12},h_{34}|X)|^2+
    C_{hh_1h_2}\tilde{C}_{hh_3h_4}|G^{(e)}_{1,0}(h,h_{12},h_{34}|X)|^2\,,
\ee
where even and odd parts of holomorphic conformal block 
read respectively
\be
    G_{1,0}^{(e)}(h,h_{12},h_{34}|X)=X^{h-\frac{1}{2}}{}_2F_1\left(h_{34}-\frac{1}{2}+h,-h_{12}+h,2h\Big|X\right)\,,
\ee
\be
    G_{1,0}^{(o)}(h,h_{12},h_{34}|X)=
    \frac{h-h_{34}}{2h}G^{(e)}_{1,0}\left(h+\frac{1}{2},h_{12},h_{34}\Big|X\right)\,.
\ee
To compute the component $g^{(0,1)}_h$ of the superblock we set
\be
\label{g01pointsfix}
    \theta_1=\theta_2=\theta_4=0\,,\quad
    \bar{\theta}_1=\bar{\theta}_2=\bar{\theta}_4=0\,.
\ee
Then superconformal partial wave evaluates to
\be
\begin{split}   \Psi^{h_1,\dots,h_4}_h&(z_1,\dots,Z_3,z_4;\bar{z}_1,\dots,\bar{Z}_3,\bar{z}_4)=
\Psi^{h_1,\dots,h_4}_h(z_1,\dots,z_4;\bar{z}_1,\dots,\bar{z}_4)\\
&+\eta^{\prime}\bar{\eta}^{\prime} \mathcal{L}_{h_1,h_2,h_3,h_4}
(h^{*}+h_{34})^2 C_{hh_1h_2}C^{*}_{hh_3h_4}\hgInt\left(h,h_{12},h_{34}+\frac{1}{2}\Big|X,\bar{X}\right)\\
    &+\eta^{\prime}\bar{\eta}^{\prime}
    \mathcal{L}_{h_1,h_2,h_3,h_4}\tilde{C}_{hh_1h_2}\tilde{C}^{*}_{hh_3h_4}
  \hgInt\left(h+\frac{1}{2},h_{12},h_{34}+\frac{1}{2}\Big|X,\bar{X}\right)\,,
\end{split}
\ee
which leads to the following expression for the function $g^{(0,1)}_{h}$:
\be
    g^{(0,1)}_h(X,\bar{X})=\tilde{C}_{hh_1h_2}C_{hh_3h_4}|G^{(o)}_{0,1}(h,h_{12},h_{34}|X)|^2+
    C_{hh_1h_2}\tilde{C}_{hh_3h_4}|G^{(e)}_{0,1}(h,h_{12},h_{34}|X)|^2\,,
\ee
with even and odd components of the conformal block given by
\be
    G_{0,1}^{(e)}(h,h_{12},h_{34}|X)=X^h{}_2F_1\left(h_{34}-\frac{1}{2}+h,-h_{12}+h,2h\Big|X\right)\,,
\ee
\be
    G_{0,1}^{(o)}(h,h_{12},h_{34}|X)=
    \frac{h+h_{34}}{2h}G^{(e)}_{0,1}\left(h+\frac{1}{2},h_{12},h_{34}\Big|X\right)\,.
\ee

\section{Integral representation of $\mathfrak{sl}(2)$ two-point torus conformal block}  \label{app:sl2twopoint}
In this section, we briefly describe some integrals that arise in shadow formalism, which yield global $\mathfrak{sl}(2)$ two-point conformal blocks. We define the integral

\begin{equation} \label{ap:2ptgcb1}
\begin{split}
  & \mathbf{\Psi}_{\Delta}^{h_1, h_2, h_3, h_4}(z_1, z_2, z_3, z_4)  =  \int_{z_4}^{z_3} dw v_{h_1, h_2, \Delta}(z_1, z_2, w) v_{1-\Delta, h_3, h_4} (w, z_3, z_4 )\\&= \int_{z_4}^{z_3} dw \frac{z_{12}^{-h_1-h_2+\Delta} z_{34}^{-h_3-h_4+1-\Delta} }{ (w-z_1)^{\Delta+h_{12} } (w-z_2)^{\Delta-h_{12} } (w-z_3)^{1-\Delta+h_{34}} (w-z_4)^{1-\Delta -h_{34} }}.
\end{split}
\end{equation}
By performing the change of variables (see, \cite{Rosenhaus2019}),
\begin{equation}
    w\to  \frac{z_1 z_{z4} w+z_4 z_{12}}{z_{24} w+z_{12}},
\end{equation}
the integral (\ref{ap:2ptgcb1})  becomes
\begin{equation} \label{ap:2ptgcb2}
\begin{split}
   \mathbf{\Psi}_{\Delta}^{h_1, h_2, h_3, h_4}(z_1, z_2, z_3, z_4)& = \frac{1}{z_{12}^{h_1+h_2} z_{34} ^{h_3+h_4}} \left( \frac{z_{24}}{z_{14}}\right)^{h_{12}}  \left( \frac{z_{14}}{z_{13}}\right)^{h_{34}} \\& \times \int_0^{\chi_1} dw \frac{\chi_1^{1-\Delta} }{ w^{1-\Delta-h_{34}} (1-w)^{\Delta-h_{12} } (\chi_1-w)^{1-\Delta+h_{34}}},
\end{split}
\end{equation}
where $\chi_1=(z_{12}z_{34})/(z_{13} z_{24})$, and it is assumed that $\chi_1<1$, hence $z_1>z_2>z_3>z_4$. By taking (\ref{ap:2ptgcb2}) one finds
\begin{equation} \label{ap:2ptgcb2-2}
\begin{aligned}
  & \mathbf{\Psi}_{\Delta}^{h_1, h_2, h_3, h_4}(z_1, z_2, z_3, z_4)\\&=
   \frac{\alpha_0 (\Delta, h_{34})}{z_{12}^{h_1+h_2} z_{34} ^{h_3+h_4}} \left( \frac{z_{24}}{z_{14}}\right)^{h_{12}}  \left( \frac{z_{14}}{z_{13}}\right)^{h_{34}}  \chi_1^{\Delta}{}_2F_1(\Delta-h_{12}, \Delta+h_{34}, 2 \Delta,\chi_1),
\end{aligned}
\end{equation}
where 
\begin{equation}
\alpha_0(\Delta, h)=\frac{\pi \csc \left(\pi  \left(\Delta +h\right)\right) \Gamma \left(\Delta -h\right)}{\Gamma (2 \Delta ) \Gamma \left(-\Delta -h+1\right)}.
\end{equation}

The expression obtained from shadow formalism \cite{Alkalaev2023} that reproduces $\mathfrak{sl}(2)$ two-point torus conformal blocks (up to an overall factor) is given by 
\begin{equation}  \label{ap:2ptgcb3}
    F_{\Delta_1, \Delta_2}^{h_1, h_2}(q, z_1, z_2)= q^{\Delta_1}\int_{\mathbf{C}_1} dw_1  \int_{\mathbf{C}_2}dw_2   v_{1-\Delta_1, h_1, \Delta_2}(w_1, z_1, w_2) v_{1-\Delta_2, h_2, \Delta_1} (w_2, z_2, w_1 q ),
\end{equation}
where the integration domains $\mathbf{C_2}, \mathbf{C}_2$ are given by (\ref{torusshadow5}). The integration over $w_2$ clearly has the form (\ref{ap:2ptgcb1}), and hence we can write 
\begin{equation}  \label{ap:2ptgcb4}
    F_{\Delta_1, \Delta_2}^{h_1, h_2}(q, z_1, z_2)= q^{\Delta_1}\int_{\mathbf{C}_1} dw_1     \mathbf{\Psi}_{\Delta_2}^{1-\Delta_1, h_1, h_2, \Delta_1}(w_1, z_1, z_2, w_1 q).
\end{equation}
The integration over $w_1$ is similar and also takes the form of (\ref{ap:2ptgcb1}). To see this, one needs to expand in series the hypergeometric function present in the expression for \\$ \mathbf{\Psi}_{\Delta_2}^{1-\Delta_1, h_1, h_2, \Delta_1}(w_1, z_1, z_2, w_1 q)$. After integrating over $w_1$, one obtains that
    
     \begin{equation} \label{ap:2ptgcb7}
       F_{\Delta_1, \Delta_2}^{h_1, h_2}(q, z_1, z_2)=  c_2(h_1, h_2, \Delta_1, \Delta_2) \mathcal{F}_{\Delta_1, \Delta_2}^{h_1, h_2}(q, z_1, z_2).
    \end{equation}
 where the coefficient $c_2$ is 
 \begin{equation} \label{coefficientc2}
 \begin{aligned}
& c_2(\text{\small$h_1, h_2,\Delta_1, \Delta_2$}) =      \alpha_0(\Delta_2, h_2-\Delta_1) \alpha_0(\Delta_1,h_1-\Delta_2)\\&=
\frac{\pi^2 \csc \left(\pi  (\Delta _1-\Delta _2+h_1)\right) \csc \left(\pi  (-\Delta _1+\Delta _2+h_2\right)) \Gamma \left(-h_1+\Delta _1+\Delta _2\right) }{\Gamma (2 \Delta _1) \Gamma (2 \Delta _2) \Gamma (-h_2+\Delta _1-\Delta _2+1) \Gamma (-h_1-\Delta _1+\Delta _2+1)} \\  & \times   \Gamma (-h_2+\Delta _1+\Delta _2),
\end{aligned}
 \end{equation}
 and $\mathcal{F}_{\Delta_1, \Delta_2}^{h_1, h_2}(q, z_1, z_2)$ is given by (\ref{torusshadow4}).
 
    Finally, let us recall that there exists another representation for the $\mathfrak{sl}(2)$ two-point torus conformal blocks in the necklace channel \cite{Alkalaev2017b}, namely 
   \begin{equation}\label{ap:2ptgcb10}
   \begin{aligned}
         \mathcal{F}_{\Delta_1, \Delta_2}^{h_1, h_2}(q, z_1, z_2)= & q^{\Delta_1} z_1^{\Delta_1-\Delta_2-h_1} z_2^{-\Delta_1+\Delta_2-h_2} \\ & \times \sum _{n=0}^{\infty } \sum _{m=0}^{\infty } \frac{q^n (\frac{z_2}{z_1}){}^{m-n} \tau _{m,n}(\Delta_2,h_2,\Delta_1) \tau _{n,m}(\Delta_1,h_1,\Delta_2)}{m! n! (2 \Delta_2)_m (2 \Delta_1)_n},
         \end{aligned}
    \end{equation}
    where
    \begin{equation}
    \begin{aligned}
&   \tau _{n,m}\left( a,b,c\right)=     \sum _{p=0}^{\min (m,n)} \frac{n! (m)^{(p)} (2 c+m-1)^{(p)} (-a+b+c)_{m-p} (a+b-c-m+p)_{n-p}}{p! (n-p)!},\\&
   (a)^{(m)} = \prod _{i=0}^{m-1} (a-i). 
   \end{aligned}
    \end{equation}
    This representation is useful when one is interested in expanding the conformal blocks in variables $z_2/z_1$ and $q$.






\bibliographystyle{JHEP}
\bibliography{biblio.bib}
\end{document}